\documentclass[a4paper,fleqn,usenatbib]{mnras}

\usepackage{newtxtext,newtxmath}
\usepackage[T1]{fontenc}
\usepackage{ae,aecompl}

\usepackage[pdftex]{graphicx}

\usepackage{graphicx}	
\usepackage{amsmath}	
\usepackage{amssymb}	

\usepackage{color}
\usepackage{epsfig}
\usepackage{comment}

\allowdisplaybreaks[1]

\usepackage{here}

\newcommand{\msun}{{\rm M}_\odot}

\newcommand{\zsun}{Z_\odot}

\newcommand{\cc}{{\rm cm}^{-3}}

\newcommand{\K}{{\rm K}}
\newcommand{\beq}{\begin{equation}}
\newcommand{\eeq}{\end{equation}}

\title[Effects of disk radiation spectra]
{Super-Eddington growth of black holes in the early Universe: effects of disk radiation spectra}

\author[]{Eishun Takeo$^{1}$\thanks{takeo@kusastro.kyoto-u.ac.jp},
Kohei Inayoshi$^{2}$\thanks{inayoshi@pku.edu.cn},
Ken Ohsuga$^{3}$,
Hiroyuki R. Takahashi$^{4}$ and
\newauthor
Shin Mineshige$^{1}$
\\
$^{1}$Department of Astronomy, Graduate School of Science, Kyoto University, 
Kitashirakawa, Oiwakecho, Sakyo-ku, Kyoto, 606-8502, Japan\\
$^{2}$The Kavli Institute for Astronomy and Astrophysics, Peking University, 
5 Yiheyuan Road, Haidian District, Beijing 100871, P. R. China \\
$^{3}$Center for Computational Sciences, University of Tsukuba, 1-1-1 Tennodai, Tsukuba, Ibaraki, 305-8577, Japan \\
$^{4}$Faculty of Arts and Sciences, Department of Natural Sciences, Komazawa University, 
1-23-1 Komazawa, Setagaya, Tokyo, 154-8525, Japan \\
}

\date{Accepted XXX. Received YYY; in original form ZZZ}

\pubyear{2018}

\begin{document}
\label{firstpage}
\pagerange{\pageref{firstpage}--\pageref{lastpage}}
\maketitle

\begin{abstract}
We investigate the properties of accretion flows onto a black hole (BH) with a mass of $M_{\rm BH}$
embedded in an initially uniform gas cloud with a density of $n_{\infty}$
in order to study rapid growth of BHs in the early Universe.
In previous work, the conditions required for super-Eddington accretion from outside the Bondi radius were studied 
by assuming that radiation produced at the vicinity of the central BH has a single-power-law spectrum $\nu^{-\alpha}$ 
at $h\nu \geq 13.6~{\rm eV}$ ($\alpha \sim 1.5$).
However, radiation spectra surely depends on the BH mass and accretion rate, and determine the efficiency of radiative feedback.
Here, we perform two-dimensional multi-frequency radiation hydrodynamical simulations taking into account more realistic 
radiation spectra associated with the properties of nuclear accretion disks.
We find that the critical density of gas surrounding the BH, above which a transitions to super-Eddington accretion occurs,
is alleviated for a wide range of masses of seed BHs ($10\lesssim M_{\rm BH}/\msun \lesssim 10^6$)
because photoionization for accretion disk spectra are less efficient 
than those for single-power-law spectra with $1\lesssim \alpha \lesssim 3$.
For disk spectra, the transition to super-Eddington is more likely to occur for lower BH masses 
because the radiation spectra become too hard to ionize the gas.
Even when accretion flows are exposed to anisotropic radiation, the effect due to radiation spectra shrinks the ionized region
and likely leads to the transition to a wholly neutral accretion phase.
Finally, by generalizing our simulation results, we construct a new analytical criterion required for super-Eddington accretion;  
$(M_{\rm BH}/10^5~\msun) (n_{\infty}/10^4~\cc) \gtrsim 2.4~ (\langle\epsilon\rangle /100~{\rm eV})^{-5/9}$,
where $\langle\epsilon\rangle$ is the mean energy of ionizing radiation from the central BH.
\end{abstract}


\section{Introduction}

Observations of bright quasars led by accreting supermassive black holes (SMBHs) 
with masses of $\gtrsim 10^9~\msun$ at high redshift $z \gtrsim 6$ (or $\lesssim 1~{\rm Gyr}$ from the Big Bang)
require rapid growth of black holes (BHs) in the early Universe \citep[e.g.][]{Fan_2004,Mortlock_2011,Wu_2015,Banados_2018}.
SMBHs are expected to play crucial roles on the history of the Universe
such as via co-evolution with their host galaxies 
\citep[e.g.][]{Silk_Rees_1998,King_2003,Murray_2005, Kormendy_Ho_2013}, 
but their formation processes are still unclear.

A possible origin of high-$z$ SMBHs is highly-accreting stellar-mass BH seeds with $\sim 100~\msun$
\citep[e.g.,][]{MadauRees01,HaimanLoeb01,VHM03, Li_2007, Alvarez_2009, alexander_natarajan_2014},
which are remnants of massive Population III stars (Pop III) 
\citep[e.g.,][]{yoshida_2008, hosokawa_2011, stacy_2012, hirano_2014, hosokawa_2016}.
Accreting gas forms an accretion disk which radiates with a luminosity of $L=\eta \dot{M}c^2$,
where $\eta$ is the radiation efficiency, $\dot{M}$ is the accretion rate, and $c$ is the speed of light.
For a rapidly accreting BH, the radiation luminosity would exceed the Eddington value $L_{\rm Edd}$,
above which the radiation force due to electron scattering overcomes the BH gravity.
Thus, the accretion rate could be limited at $\dot{M} \leq L_{\rm Edd}/(\eta c^2)$.
As result of this, the BH growth timescale from light seeds becomes significantly longer than the age of the Universe when 
high-$z$ SMBHs already exist ($\gtrsim 1~{\rm Gyr}$) in cases with $\eta \sim 0.1$ \citep{Soltan_1982,Yu_Tremaine_2002}.
Thus, rapid growth of BHs via super-Eddington gas accretion is an attractive pathway to high-$z$ SMBHs.

Another possibility is more massive BH seeds with $\sim 10^3-10^5~\msun$
formed by direct collapse of supermassive stars in protogalaxies
\citep[e.g.,][]{loeb_rasio_1994, oh_haiman_2002, bromm_loeb_2003, begelman_2006, 
regan_haehnelt_2009a, regan_haehnelt_2009b, hosokawa_2012, hosokawa_2013,  
IOT14, 2014MNRAS.445.1056V,  IT_2015,  IVK_2015,  Chon_2016,  regan_2016a, 
regan_2016b,hirano_2017,inayoshi_2018}
and runaway stellar collisions
\citep[e.g.,][]{omukai_2008, 2009ApJ...694..302D,2015MNRAS.451.2352K,2016MNRAS.457.2423Y,
Stone_2017,Sakurai_2017,Reinoso_2018}.
Even for such heavy seeds, we need to require a high duty cycle of BH growth at the Eddington accretion rate.

The possibility of super-Eddington accretion has been explored by many authors.
In fact, some of ultra-luminous X-ray sources are considered to be stellar-mass BHs accreting at super-Eddington rates 
\citep{Fabbiano+1989,King+2001, Watarai+2001}, and narrow-line Seyfert-1 galaxies are presumably super-Eddington accretors
\citep[e.g.,][]{Wang+1999, Mineshige+2000}.
%
By means of two-dimensional radiation hydrodynamical simulation, 
\cite{ohsuga+05} have revealed that super-critical accretion is realized
as long as sufficient gas is supplied at the vicinity of the central BH.
\cite{ohsuga_mineshige_2007} have concluded that trapping of diffusive photons in the optically-thick accretion disk 
and anisotropic radiation are crucial to realize super-Eddington accretion
\citep[see also][]{Begelman_1979,ohsuga+2009,ohsuga_mineshige2011,Jiang_2014,Sadowski_2015,takahashi2016,kitaki_2018}.
On the other hand, gas supply from larger scale can be significantly suppressed due to photoionization heating and radiation momentum
\citep[e.g.,][]{Ciotti_Ostriker_2001,Milosavljevic_2009a,Milosavljevic_2009b,Alvarez_2009,
2009ApJ...699...89C,Park_Ricotti_2011,Park_Ricotti_2012}.
Since the connection between BH feeding and feedback has been understood poorly yet,
previous works with semi-analytical models adopted various prescriptions for BH accretion rates
in the assembly history of dark matter halos
\citep[e.g.][]{VR_2005,TH09,Alexander_2014,Madau_2014,2015MNRAS.448..104P,Valiante_2016,Pezzulli_2016, pezzulli_2017, valiante_2018}.

Recently, \cite{inayoshi+16} found the conditions for super-critical accretion in a spherically symmetric system 
exposed to intense radiation from the BH with $L\simeq L_{\rm Edd}$.
When the size of an ionized region $r_{\rm H_{II}}$ surrounding the accreting BH is smaller than the Bondi radius $r_{\rm B}$,
the ionized region collapses due to intense inflows of neutral gas and thus the accretion system transits 
to an isothermal ($\approx 8000~\K$) Bondi accretion solution with a high accretion rate of $\gtrsim 5000~L_{\rm Edd}/c^2$.
The transition criterion is written as,
\begin{equation}
M_{\rm BH} \times {n_\infty} \ga 10^9~\msun~\cc ~ (T_\infty/10^4~\K)^{3/2},
\label{eq:super_Edd}
\end{equation}
where $n_\infty$ and $T_\infty$ are density and temperature of the ambient gas.
Even with super-Eddington radiation feedback ($L>L_{\rm Edd}$), the above criterion does not change significantly \citep{Sakurai_2016}.
Moreover, the criterion for super-critical accretion is alleviated under anisotropic radiation fields \citep{sugimura_2017,takeo_2018}.
We note that the transition criterion is characterized by a quantity of $M_{\rm BH} \times {n_\infty}$
because $r_{\rm H_{II}}/r_{\rm B} \propto (M_{\rm BH} \times {n_\infty})^{-2/3}$.
Here $r_{\rm H_{II}}$ is estimated by the Str\"{o}mgren radius 
\begin{equation}
r_{\rm Strm} \equiv \left(\dfrac{3\dot{N}_{\rm ion}}{4\pi n_{\rm H_{II}}^2 \alpha_{\rm B}}\right)^{1/3},
\label{strm}
\end{equation} 
where $\dot{N}_{\rm ion}$ is the emission rate of ionizing photons, 
$n_{\rm H_{II}}$ is the number density of ionized gas, 
and $\alpha_{\rm B}$ is the case B radiative recombination rate.

The previous work assumed that radiation from the central region has a single-power-law (hereafter PL) spectrum 
with $L_{\nu} \propto \nu^{-\alpha}$,
where the spectral index $\alpha$ is often set to 1.5
\footnote{
Observed and calculated radiation spectra at $13.6\la h\nu/{\rm eV} \la 2~{\rm keV}$ are fit 
by power-law with $\alpha \sim 1.1 - 1.5$ \citep[][and references therein]{liu_2003}.
}
(see also \S\ref{sec:discuss}).
However, the shape of the disk radiation spectrum would be more complicated.
According to the analytic solutions of accretion disks,
the disk surface temperature is described as $T_{\rm eff}(R) \propto R^{-p}$ \citep[e.g.,][]{mineshige+1994,kato+2008},
where $R$ is the distance from the central BH.
Assuming the disk surface locally emits blackbody radiation with $T_{\rm eff}(R)$, 
the spectrum is written as $L_{\nu} \propto \nu^{3-(2/p)}$;
$L_{\nu} \propto \nu^{1/3}$ ($p=3/4$) in the standard disk case \citep{SS_1973}, 
and $L_{\nu} \propto \nu^{-1}$ ($p=1/2$) in the slim disk case \citep{abramowicz_1988}
(see \S\ref{sec:model} for more details). 
Moreover, since the disk temperature reaches $\sim 10^7~\K$, the maximum energy of the continuum spectrum 
is as high as $\sim 1~{\rm keV}$ \citep[e.g.,][]{watarai_2006},
which is much higher than the mean photon energy of the PL spectrum $\approx 40.8~{\rm eV}$.
This fact implies that radiative feedback effects for disk spectra are less efficient than that for the PL spectrum
because the cross section to bound-free absorption of hydrogen atoms is 
$\sigma_{\rm bf,H} \propto \nu^{-3}$ \citep[e.g.,][]{draine_2011}.
On the other hand, electrons primarily produced by X-ray ionization are energetic enough to ionize the ambient gas
\citep[e.g.][]{shull_1979, shull_van_1985, ricotti_2002}.

In this paper, we investigate the conditions for super-Eddington accretion under radiation with disk spectra
associated with the standard and slim accretion disk model.
We performed two-dimensional hydrodynamical simulations, including one-dimensional multi-frequency radiation transfer
and primordial chemical reaction networks.
We first conduct simulations under isotropic radiation from the central accretion disk.
We construct an analytical formula for the criterion required for super-Eddington accretion under isotropic radiation 
and show that the conditions are alleviated for a wide range of BH masses, compared to the cases with single-PL spectra.
Next, we perform simulations of accretion flows exposed to anisotropic radiation
and investigate effects of the disk spectrum onto the inflow rate 
and conditions for the transition to a wholly neutral accretion phase.

The rest of this paper is organized as follows.
In Section \ref{sec:2}, we describe the methodology of our numerical simulations.
In Section \ref{sec:resl}, we show our simulation results and give the conditions required for super-Eddington accretion.
In Section \ref{sec:discuss}, we discuss the analytical formula for the transition, the stability of highly-accreting system 
after the transition, and caveats of our simulation setups.
In Section \ref{sec:conc}, we summarize the conclusion of this paper.

\section{Methods}
\label{sec:2}

Our goal is to study conditions for super-Eddington accretion led by gas supply from larger scales.
Gas accretion begins from a critical radius, the so-called Bondi radius, defined by
\begin{equation}
r_{\rm B}\equiv \frac{GM_{\rm BH}}{c_{\infty}^2} \simeq 1.97\times10^{14}
~m_{\rm BH}
~T_{\infty,4}^{-1} \ {\rm cm},
\label{eq:bondi_radius}
\end{equation}
and the Bondi accretion rate for isothermal gas is given by
\begin{equation}
  \dot{M}_{\rm B} \equiv \pi e^{3/2} \rho_\infty \frac{G^2 M_{\rm BH}^2}{c_{\infty}^3},
  \label{bondi_rate}
\end{equation}
where $m_{\rm BH} \equiv M_{\rm BH}/\msun$, 
$T_{\infty,4} \equiv (T_\infty/10^4~\K)$,
$c_{\infty} \equiv \sqrt{\gamma \mathcal{R} T_{\infty} /\bar{\mu}}$ is the sound speed,
$\gamma$ is the specific heat ratio, 
$\mathcal{R}$ is the gas constant, and
$\bar{\mu}$ is the mean molecular weight.
Note that the Bondi radius and rate as reference values are calculated by setting 
$\gamma=1$, $\bar{\mu}=1.23$ and $T_{\infty}=10^4 {\rm K}$.

\subsection{The code}
We perform two-dimensional hydrodynamical simulations of axisymmetric flows
with one-dimensional radiation transfer and chemical reaction networks \citep{takeo_2018}.
Here we adopt the hydrodynamical simulation code developed in \cite{takahashi_ohsuga_2013}.
The advection terms for the ideal fluid are computed using the Harten-Lax-vanLeer Riemann solver \citep{harten_1983}, and 
the second order accuracy in space and time are ensured \citep{van_leer_1977}.
We adopt the spherical coordinates of $(r, \theta, \phi)$ 
with the polar axis ($\theta = 0$ and $\pi$) perpendicular to the disk plane.
We add the radiation and chemical codes taken from \cite{inayoshi+16} with necessary modifications.

\subsection{Basic equations}
\label{sec:BE}
The basic equations of the hydrodynamical part are the equation of continuity
\begin{equation}
  \frac{\partial \rho}{\partial t} + \nabla\cdot(\rho {\boldsymbol v}) = 0,
  \label{renzoku}
\end{equation}
the equations of motion
\begin{equation}
  \frac{\partial \left(\rho v_r\right)}{\partial t} + \nabla\cdot(\rho v_r {\boldsymbol v}) 
  = -\frac{\partial p}{\partial r} + \rho\left( \frac{v_{\theta}^2}{r} +\frac{v_{\phi}^2}{r} \right) - \rho \frac{\partial \psi}{\partial r}+ f_{\rm rad},
  \label{eomr}
\end{equation}
\begin{equation}
  \frac{\partial \left(\rho rv_{\theta}\right)}{\partial t} + \nabla\cdot(\rho r v_{\theta} {\boldsymbol v}) 
  = -\frac{\partial p}{\partial \theta} + \rho v_{\phi}^2\cot{\theta},
  \label{eomt}
\end{equation}
\begin{equation}
  \frac{\partial \left(\rho rv_{\phi}\sin{\theta}\right)}{\partial t} + \nabla\cdot\left(\rho r v_{\phi}\sin{\theta} {\boldsymbol v}\right) = 0,
  \label{eomp}
\end{equation}
and the energy equation
\begin{equation}
  \frac{\partial e}{\partial t} + \nabla\cdot[(e+p){\boldsymbol v}] = -\frac{GM_{\rm BH}\rho}{r^2}v_r -\Lambda + \Gamma .
  \label{eneeq}
\end{equation}
where $\rho$ is the gas density, ${\boldsymbol v}=(v_r,v_{\theta}, v_{\phi})$ is the velocity, 
$p$ is the gas pressure, and $f_{\rm rad}$ is the radiation force.
We consider the gravity of the central BH ($r=0$) and neglect the gas self-gravity. 
Since the general relativistic effect is negligible, the gravitational potential is given by 
$\psi=-GM_{\rm BH}/r$.
The total energy per volume is defined as $e \equiv e_{\rm int}+\rho |{\boldsymbol v}|^2/2$, 
$e_{\rm int}$ is the gas internal energy density, 
$\Lambda$ is the cooling rate per volume, and $\Gamma$ is the radiative heating rate.
We assume the equation of state of ideal gas as $p=(\gamma-1)e_{\rm int}$ for $\gamma=5/3$.

We solve the multi-frequency radiative transfer equation
\begin{equation}
\frac{1}{r^2}\frac{{\rm d}}{{\rm d}r}(r^2F_{\nu}) = -\rho\kappa_{\nu}cE_{\nu},
\label{rteq}
\end{equation}
where $F_{\nu}$ is the radiation flux, $E_{\nu}$ is the radiation energy density, and 
$\kappa_{\nu}$ is the absorption opacity.
The radiation field is assumed to be steady
because the light crossing time is much shorter than the hydrodynamical timescale.
The frequency range is set to $h\nu_{\rm min}(=13.6~{\rm eV}) \leq h\nu \leq h\nu_{\rm max}(=100~{\rm keV})$,
where $h$ is the Planck constant.
We note that only the radial component of the radiation flux is calculated because non-radial components produced by
radiative recombination is negligible (see \S4 in \citealt{takeo_2018}).
Since the ionized gas is optically thin to electron scattering,
we assume $F_{\nu} = cE_{\nu}$ on the right-hand-side of Eq. (\ref{rteq}).

We consider cooling processes associated with ${\rm H, He, He^+}$ atoms and free-free emission \citep{glover+07}, 
assuming the optically-thin cooling rates.
In order to estimate their rates, we solve chemical reaction networks including six species of 
$\rm H, H^+, He, He^+, He^{++}$, and ${\rm e^-}$. 
The abundance of He nuclei relative to H nuclei is set to $8.33\times10^{-2}$.
Here we consider photoionization, collisional ionization and radiative recombination \citep{abel+97,glover+07},
including effects of the secondary ionization (see below). 
Photoionization due to diffusive recombination photons is neglected, i.e.,
the case B recombination rate is adopted instead of the case A rate. 
The cooling/heating term in the energy equation (Eq. \ref{eneeq}),
the chemical reaction, and the radiative transfer equation (Eq. \ref{rteq})
are updated with an implicit method
in order to 
solve them stably and save computation time.
We set the time steps by setting the Courant number to 0.4.

The ionization rate coefficients and photoionization heating rates 
are calculated with the photon-conserving method \citep{Whalen_2006}.
The primary ionization rates $k_{{\rm ph},i}^{\rm p}$ ($i={\rm H, He}$, and ${\rm He^+}$) are estimated as
\begin{equation}
k_{{\rm ph},i}^{\rm p} = \int_{\nu_{\rm min}}^{\nu_{\rm max}} {\rm d}\nu~\dfrac{F_{\nu}}{h\nu}~ \sigma_{{\rm bf},i},
\end{equation}
where $\sigma_{{\rm bf},i}$ is the bound-free cross section.
Since the energy of electrons produced by primary ionization is higher than the ionization potential energy,
the electrons further ionize neutral hydrogen nearby \citep[e.g.][]{shull_1979, shull_van_1985}.
The secondary ionization rates for species $j={\rm H}$, and ${\rm He}$ are
\begin{equation}
k_{{\rm ph},j}^{\rm s} = \sum_{i={\rm H, He}} \int_{\nu_{\rm min}}^{\nu_{\rm max}} {\rm d}\nu~\dfrac{F_{\nu}}{h\nu}
~\sigma_{{\rm bf}, j}  ~\Phi^{j}(E_{i},x_{\rm H^+}) ~\dfrac{x_{i}}{x_{j}},
\end{equation}
where $x_j$ is the abundance of species $j$, $\Phi^{j}(E_{i},x_{\rm H^+})$ is the fraction of secondary ionization of species $j$ 
per primary electron of energy $E_{i}\equiv h\nu - I_{i}$, 
and $I_i$ is the ground state ionization potential energy of the species $i$.
The total photoionization rate is given by the sum of primary and secondary ionization rates.
The photoionization heating rate ($i={\rm H},~{\rm He}$, and ${\rm He^+}$) is 
\begin{equation}
\Gamma_{i} = \int_{\nu_{\rm min}}^{\nu_{\rm max}} {\rm d}\nu~\dfrac{F_{\nu}}{h\nu} ~\sigma_{{\rm bf},i} ~E_h(E_{i},x_{\rm H^+}), 
\end{equation}
$E_h$ is the energy of primary electrons deposited as heat.
We adopt the functional forms of $\Phi^{\rm H}$, $\Phi^{\rm He}$, and $E_h$ \citep{ricotti_2002}.
Note that secondary ionization of ${\rm He}^{+}$ is negligible \citep{shull_van_1985}.
The radiation force caused by electron scattering and bound-free absorption is given by
\begin{equation}
  f_{\rm rad} = \frac{nx_{\rm e}}{c}\int_{\nu_{\rm min}}^{\nu_{\rm max}}\sigma_{\rm es}F_{\nu}{\rm d}\nu + \frac{\Gamma^{\rm p}}{c},
  \label{f_rad}
\end{equation}
where $\Gamma^{\rm p}$ is the sum of the heating rates due to primary ionization of ${\rm H,~He}$, and ${\rm He^+}$ atoms.

\subsection{Disk radiation models}
\label{sec:model}

In order to study the effect of radiation produced from the nuclear accretion disk,
we adopt models for the radiation spectra.
Since the accretion rate through the disk we consider is as high as $\dot{m}\equiv \dot{M}/\dot{M}_{\rm Edd}\ga 10^{-2}$,
where $\dot{M}_{\rm Edd}\equiv L_{\rm Edd}/c^2$ and $L_{\rm Edd}$ is the Eddington luminosity,
the disk emission can be approximated as multi-color blackbody spectra \citep[e.g.,][]{kato+2008}.
The specific radiation luminosity is calculated as 
\begin{equation}
L_{\nu}=2\int_{R_{\rm in}}^{R_{\rm out}}~{\rm d}R~2\pi R ~B_{\nu}[T_{\rm eff}(R)],
\end{equation} 
where $B_{\nu}(T_{\rm eff})$ is the Black body intensity with an effective temperature of $T_{\rm eff}$, 
and $R_{\rm in(out)}$ is the inner (outer) radius of the disk ($R$ is the radius of the cylindrical coordinate).
The disk outer radius is set to $R_{\rm out}=10^4~r_{\rm Sch}$ as our fiducial value,
where $r_{\rm Sch} \equiv 2GM_{\rm BH}/c^2$ is the Schwarzschild radius.
Note that we discuss the dependence of our results on the choice of $R_{\rm out}$ in \S\ref{sec:discuss}.
For $\dot{m}  < 10$, we set the inner disk radius to the inner most stable circular orbit (ISCO) for 
a non-spinning BH ($R_{\rm in}=3~r_{\rm Sch}$).
In the slim disk cases with $\dot{m} > 40$, we set $R_{\rm in}=1.1~r_{\rm Sch}$
because the gas is optically thick even inside the ISCO.
For $10 \leq \dot{m} \leq 40$, we estimate $R_{\rm in}$ with a linear interpolation in the plane of $\log \dot{m}-\log R_{\rm in}$.
The radial structure of the effective temperature is given by \cite{watarai_2006} as
\begin{equation}
T_{\rm eff}(R) = 2.5 \times 10^7~\K~ f^{1/8} \left(\dfrac{m_{\rm BH}}{10}\right)^{-1/4} \left(\frac{R}{r_{\rm sch}}\right)^{-1/2} \mathcal{F}(R,~\dot{m}),
\label{disk_sol}
\end{equation}
where
\begin{equation}
\mathcal{F}(R,~\dot{m}) \equiv 
\begin{cases}
\left(1-\sqrt{3r_{\rm Sch}/R}\right)^{1/4} & {\rm for}~ \dot{m}\leq10,\\
\left(1-\sqrt{R_{\rm in}/R}\right)^{1/4} & {\rm for}~ 10<\dot{m}<40,\\
1 & {\rm for}~\dot{m}\geq 40,
\end{cases}
\end{equation}
and $f$ is a function of $R$ and $\dot{m}$ 
which connects the standard and slim disk solution smoothly\footnote{
The function $f$ is defined as a ratio of the advection cooling rate to the viscous heating rate at radius $R$
and satisfies
\begin{equation}
f = 0.5\left( D^2X^2 + 2 -DX \sqrt{D^2X^2 + 4} \right),
\end{equation}
where $X \equiv R/(\dot{m} r_{\rm Sch})$, and we set $D=2.81$.}.
When the accretion rate is sufficiently high ($\dot{m}\gg 1$), the advection cooling timescale is shorter than 
the photon diffusion timescale within a characteristic radius, so-called the photon-trapping radius 
$R_{\rm tr} \equiv \dot{m}r_{\rm Sch}$, where $f\simeq 1$ and $\mathcal{F}\simeq 1$.
At $R \ga R_{\rm tr}$, optically-thick radiative cooling in the disk is dominant and $f \propto R^{-2}$.
Thus, most of the radiation is produced within the trapping radius, and the bolometric luminosity 
is expressed as
\begin{equation}
\frac{L}{L_{\rm Edd}} = 2\left[1+\ln \left(\frac{\dot{m}}{20}\right)\right],
\label{eq:Lmdot}
\end{equation}
for $\dot{m}>20$ \citep{watarai+00}.
We also model the angular dependence of radiation fields in the same way as in 
\cite{takeo_2018},
\begin{equation}
  F_{\nu}(r=r_{\rm min},\theta) = \frac{\displaystyle ({\mathcal N}+1)L_{\nu}}
  {\displaystyle 4\pi r_{\rm min}^2}\cos^{\mathcal N}{\theta},
  \label{frad_in}
\end{equation}
where $r_{\rm min}$ is the size of the inner-most grid (see \S\ref{sec:IBC}) and $\mathcal{N}$ characterizes
the anisotropy of radiation fields.
In this study, we explore both isotropic cases ($\mathcal{N}=0$) and anisotropic cases ($\mathcal{N}=4$).

%
Furthermore, we consider the photon redshift effect, i.e.,
the radiation intensity observed at infinity $I_{\nu_{\rm obs}}$ is connected with 
the intensity $I_{\nu_{\rm em}}$ at the photon emitting point ($R=R_{\rm em}$) on the disk surface
as $I_{\nu_{\rm obs}} = (\nu_{\rm obs}/\nu_{\rm em})^3 I_{\nu_{\rm em}}$,
where we assume $\nu_{\rm obs}/\nu_{\rm em} = (1-r_{\rm Sch}/R)^{1/2}$
for simplicity
\footnote{
Although general relativistic (GR) effects on accretion discs are studied recently
\citep[e.g.,][]{Sadowski_2015,takahashi2016, Narayan+2017},
we do not adopt the details of their findings in our disk model.
In fact, GR effects would not be so important for super-Eddington accretion discs 
because the photosphere $r_{\rm ph} \sim \dot{m}r_{\rm g}$ becomes 
much larger than the Schwarzschild radius \citep[e.g.,][]{kitaki+2017}.
}.

\begin{figure}
\begin{center}
\includegraphics[width=8.3cm]{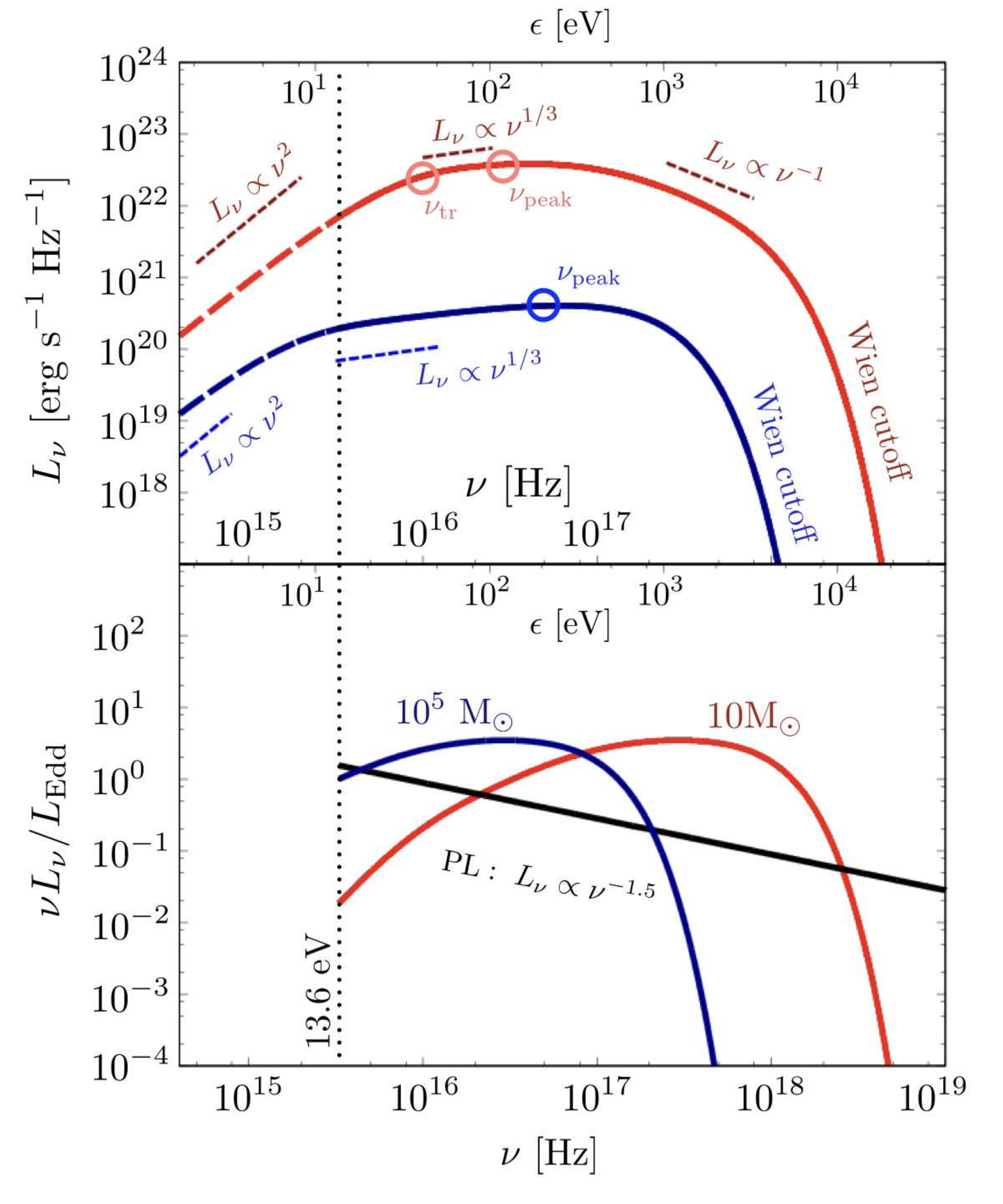}
\end{center}
\vspace{-4mm}
\caption{
Spectral models for radiation emitted from the nuclear BH accretion disk.
In the top panel, disk spectra with $m_{\rm BH}=10$ 
for $\dot{m}=1$ (blue; the standard disc) and $\dot{m}=10^3$ (red; the slim disc) are shown.
Note that non-ionizing photons with lower energies of $h\nu < 13.6~{\rm eV}$ (dashed) are not taken into account.
In the bottom panel, we present disk spectra with $\dot{m}=10^3$ for  $m_{\rm BH}=10$ (red) and $m_{\rm BH}=10^5$ (blue),
and the PL spectrum $L_{\nu} \propto \nu^{-1.5}$ with $\dot{m}=10^3$ (black).
}
\label{nulnu_basic}
\end{figure}

\subsection{Emergent spectra from the disk}
\label{sec:diskmodel}

In the top panel of Fig. \ref{nulnu_basic}, we show the spectral shape of radiation from an accretion disk 
around the central BH with $m_{\rm BH}=10$ at an accretion rate of $\dot{m}=1$ (blue) and $\dot{m}=10^3$ (red).
For the standard disk case ($\dot{m}=1$), the spectrum is expressed by a multi-color Black body spectrum of
\begin{equation}
L_{\nu}^{\rm D,st} = 5.7 \times 10^{13}~ m_{\rm BH}^{4/3}~\dot{m}^{2/3}~\nu^{1/3}~{\rm erg~s^{-1}~Hz^{-1}}
\end{equation}
at the frequency range of $\nu_{\rm out}\lesssim \nu \lesssim \nu_{\rm peak}$, where
\begin{equation}
\nu_{\rm out} \equiv 3.16 \times 10^{18}\ m_{\rm BH}^{-1/4} \dot{m}^{1/4}\left(\frac{{R}_{\rm out}}{r_{\rm Sch}}\right)^{-3/4}~{\rm Hz},
\end{equation}
is the frequency of photons emitted from $R=R_{\rm out}$ and 
\begin{equation}
\nu_{\rm peak} \equiv 1.01 \times 10^{17} m_{\rm BH}^{-1/4}\dot{m}^{1/4}~{\rm Hz},
\end{equation}
is the peak frequency of the spectrum.
The spectral shape is expressed by the Rayleigh-Jeans slope ($L_\nu \propto \nu^2$) at $\nu < \nu_{\rm out}$,
and has an exponential cutoff, the so-called Wien cutoff, at $\nu > \nu_{\rm peak}$.

For the slim disk case ($\dot{m}=10^3$), the disk spectrum has an additional component
associated with the modification of the effective temperature as described in Eq. (\ref{disk_sol}),
\begin{equation}
L_{\nu}^{\rm D,slim} = 5.15 \times 10^{38}~m_{\rm BH}~\nu^{-1}~{\rm erg~s^{-1}~Hz^{-1}}.
\end{equation}
at the frequency range of $ \nu_{\rm tr} < \nu < \nu_{\rm peak}$, where 
\begin{equation}
\nu_{\rm tr}\equiv 10^{17} (m_{\rm BH}/10 )^{-1/4} (\dot{m}/10^3 )^{-1/2}~{\rm Hz},
\end{equation}
corresponds to the frequency of photons emitted from $R_{\rm tr}$.


In the bottom panel of Fig. \ref{nulnu_basic}, we compare three different radiation spectra:
a single PL with an index of $\alpha=1.5$ (black), and disk spectra with 
$M_{\rm BH}=10$ (red) and $10^5~\msun$ (blue) for $\dot{m}=10^3$.
The disk spectrum with $m_{\rm BH}=10$ is harder than the PL spectrum: in fact, 
the difference of the luminosities at $h\nu = 13.6~{\rm Hz}$ is the order of $\sim 10^2$.
The disk spectrum becomes softer as the BH mass increases following $T_{\rm eff} \propto m_{\rm BH}^{-1/4}$.

\begin{table}
\begin{center}
\caption{Model parameters and results for isotropic radiation cases.}
\begin{tabular}{ccccc}
Model & $M_{\rm BH}(\msun)$ & $n_{\infty}(\cc)$ & transition &$t_{\rm tran,end}~(t_{\rm dyn})$ \\
\hline 
1e0M18N0 & $1$ & $1\times10^8$ &  N &5.2\\
1e0M38N0 & $1$ & $3\times10^8$ &  N &5.2 \\
1e0M58N0 & $1$ & $5\times10^8$ & Y &{\bf 4.3}\\
1e0M19N0 & $1$ & $1\times10^9$ & Y &{\bf 1.8} \\
1e0M39N0 & $1$ & $3\times10^9$ & Y &{\bf 0.75} \\
&&& \\
1e1M37N0 & $10$ & $3\times10^7$ &  N &5.5  \\
1e1M18N0 & $10$ & $1\times10^8$ & Y &{\bf 2.9}  \\
&&&\\
1e2M36N0 & $10^2$ & $3\times10^6$ &  N &5.6  \\
1e2M17N0 & $10^2$ & $1\times10^7$ & Y &{\bf 4.4}  \\
1e2M37N0 & $10^2$ & $3\times10^7$ & Y &{\bf 1.6}  \\
&&&\\
1e5M14N0 & $10^5$ & $1\times10^4$ &  N &6.8  \\
1e5M34N0 & $10^5$ & $3\times10^4$ & Y &{\bf 6.1}  \\
1e5M54N0 & $10^5$ & $5\times10^4$ & Y &{\bf 2.7}  \\
1e5M15N0 & $10^5$ & $1\times10^5$ & Y &{\bf 1.3}  \\
\hline
\end{tabular}
\label{models}
\end{center}
Column (1) model ID, (2) BH mass, (3) ambient gas density, 
(4) symbols Y (N) denoting that the transition occurs (does not occur) within the simulation time, 
and (5) the time when the transition occurs $t_{\rm tran}$ (bold)
 and the duration $t_{\rm end}$ (thin, for models without transitions) in units of 
 $t_{\rm dyn}(\equiv r_{\rm B}/c_\infty)$.
\end{table}

\subsection{initial and boundary conditions}
\label{sec:IBC}

We set a computational domain of $r_{\rm min} \leq r \leq r_{\rm max}$ and $0 \leq \theta \leq \pi$, 
where $(r_{\rm min},~r_{\rm max}) = (0.007~r_{\rm B}, ~6~r_{\rm B})$ for isotropic radiation,
and $(r_{\rm min},~r_{\rm max}) = (0.07~r_{\rm B}, ~60~r_{\rm B})$ for anisotropic radiation. 
In the case with anisotropic radiation, we set a larger simulation box because the ionized region toward 
the bipolar directions tends to be larger than that in isotropic cases.
We set logarithmically-spaced grids in the radial direction 
and uniformly-spaced grids in the polar direction. 
The number of the grid points is set to $(N_r, N_{\theta}) = (100,120)$.

As our initial conditions, we set a neutral uniform and static (${\boldsymbol v}= 0$) gas cloud with a density 
$n_{\infty}$ and temperature $T_{\infty}=10^4~\K$.
The BH mass is assumed to be constant throughout the simulations.
Since our main goal is to derive the conditions for super-Eddington 
transitions,
we explore a wide range of the ambient density and BH mass:
$10^4 \leq n_{\infty}/\cc \leq 3\times 10^9$ and $1\leq M_{\rm BH}/\msun \leq 10^5$.
Our model setup with isotropic radiation is summarized in Table \ref{models}.
We impose the absorption inner-boundary conditions which damps the gas density, 
the velocity, and gas pressure smoothly \citep[e.g.][]{kato+04},
and the free outer-boundary conditions for three components of the velocity 
and the specific entropy.
We also fix the same gas density at $r=r_{\rm max}$ as the initial value for grids with an inflow velocity 
i.e., $v_r(r=r_{\rm max}) < 0$, otherwise the free boundary condition is imposed for the density.
The reflection symmetry with respect to the polar axis is imposed for 
non-radial components of the velocity.

\begin{figure}
\begin{center}
\includegraphics[width=8.5cm]{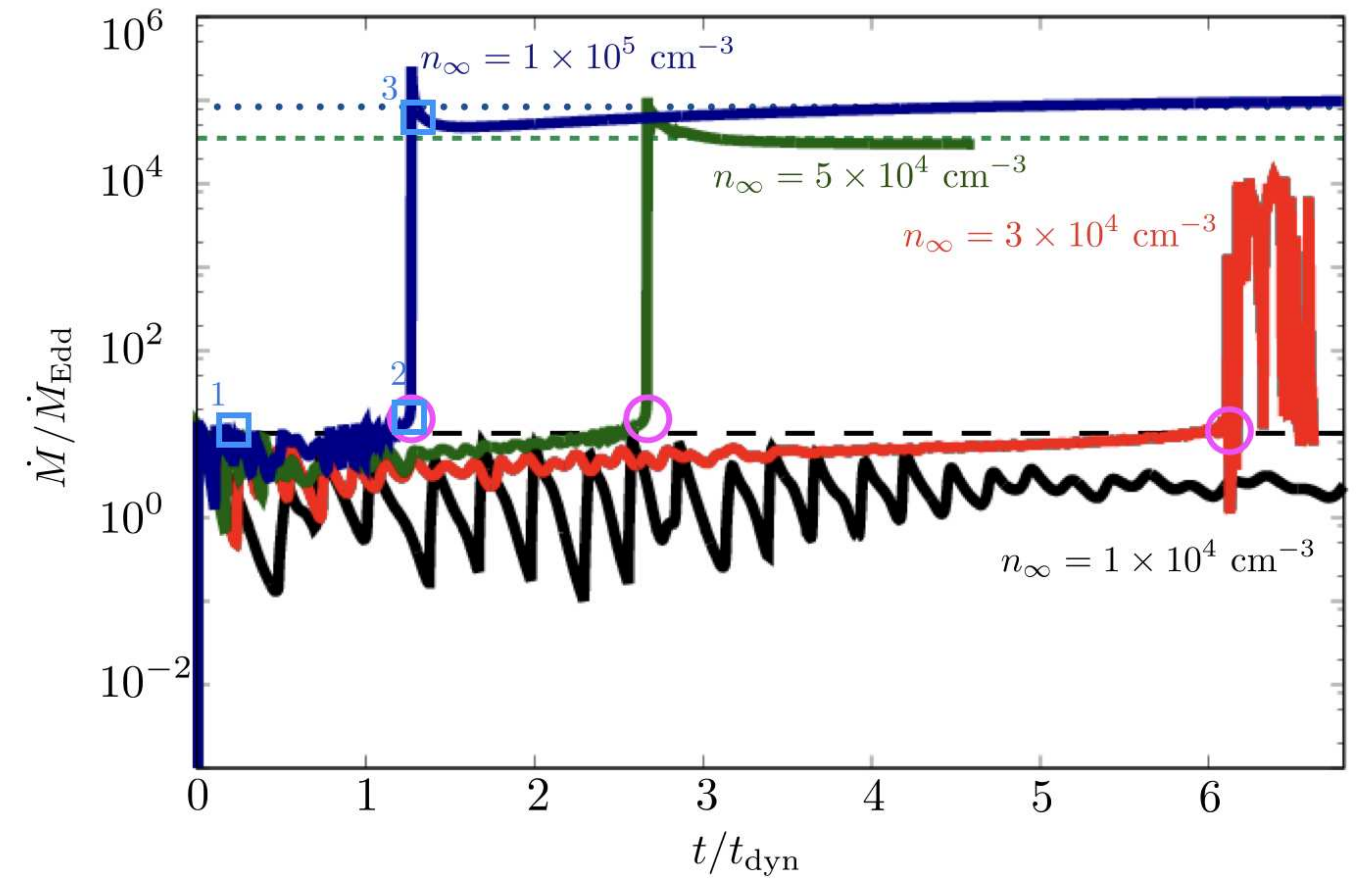}
\end{center}
\vspace{-2mm}
\caption{
Time evolution of accretion rates onto a BH with a mass of $M_{\rm BH} = 10^5~\msun$
embedded in a gas cloud with different densities of $n_{\infty} = 10^4$ (black), 
$3\times 10^4$ (red), $5\times 10^4$ (green), and $10^5~\cc$ (blue). 
For the lowest density ($n_\infty=10^4~\cc$), the gas accretion occurs episodically due to photoionization and heating. 
For higher densities with $n_\infty >3\times 10^4~\cc$, the accretion rates behave similarly in the early stages where the rates 
are limited around $\dot{M}=10~\dot{M}_{\rm Edd}$ (black dashed), but transit into rapid accretion phases where the rates are 
approximated as the Bondi accretion rates for the given ambient densities (horizontal dotted lines).
The transition epochs are marked by open circles.
Open squares indicate the epochs at which we show the radial profiles in Fig. \ref{prof}.
}
\label{t_md_1e5M}
\end{figure}

\begin{figure*}
\begin{center}
\includegraphics[width=16cm]{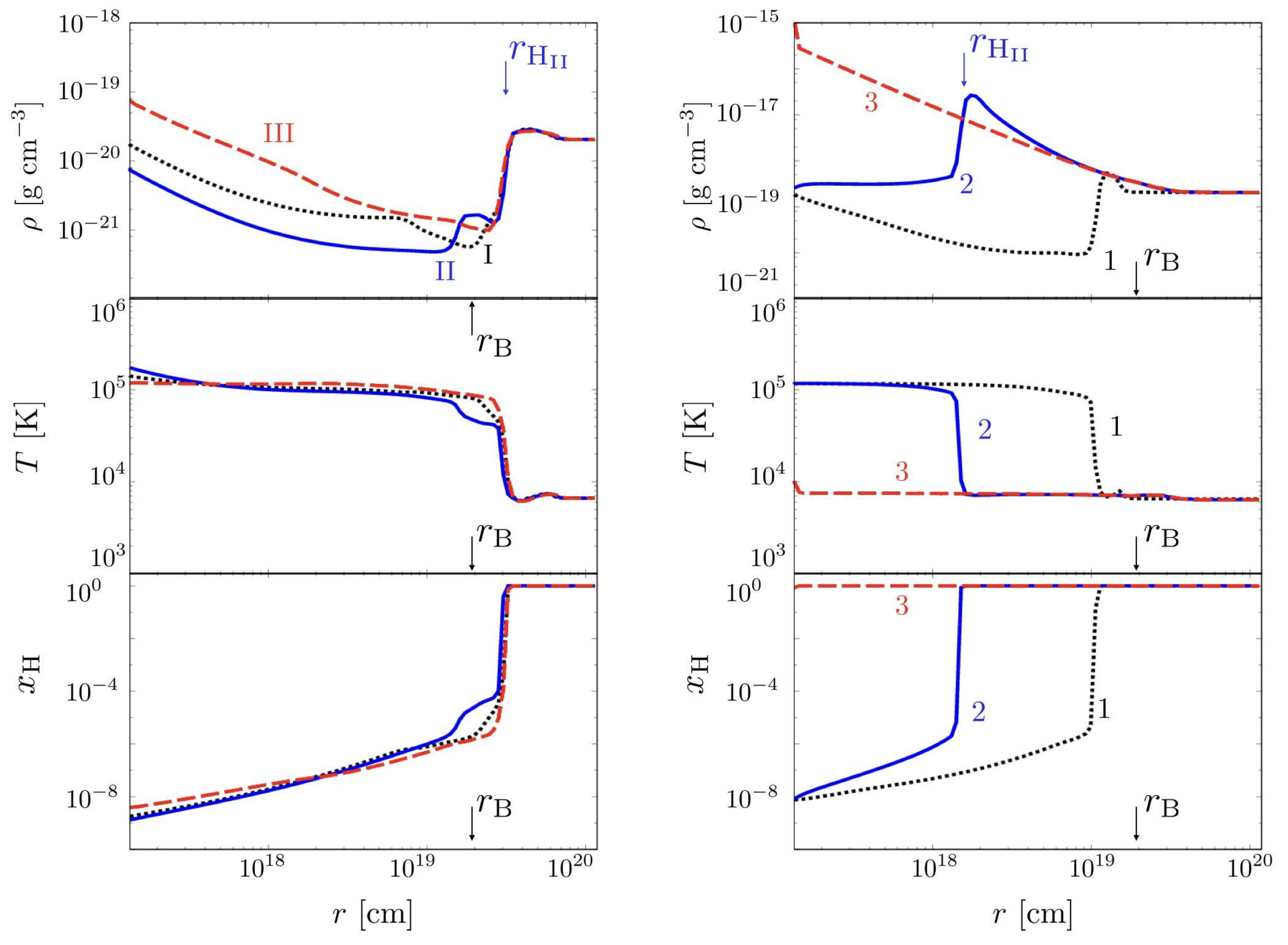}
\end{center}
\vspace{-4mm}
\caption{
Radial structure of the gas density (top), temperature (middle), neutral fraction (bottom) at the equatorial plane.
In the left panels, we present the profiles for Model 1e5M14N0, where the accretion occurs episodically without 
a transition to super-Eddington phases, at three different epoch during an oscillation: 
$t/t_{\rm dyn} = 1.46$ (dotted), $1.55$ (solid), and $1.69$ (dashed).
In the right panels, we show those for Model 1e5M15N0, where the accretion rate transits to a super-Eddington value,
at $t/t_{\rm dyn} = 0.334$ (dotted), $1.25$ (solid) and $1.38$ (dashed).
For the case without the transition, the location of the ionization front $r_{\rm H_{II}}$ is outside the Bondi radius, while 
the ionized region is always confined inside the Bondi radius for the case with the transition.
}
\label{prof}
\end{figure*}

\section{Results}
\label{sec:resl}


Fig. \ref{t_md_1e5M} presents the time evolution of accretion rates onto a BH with $M_{\rm BH} = 10^5~\msun$
for $n_{\infty} = 1\times10^4~\cc$ (black), $3\times 10^4~\cc$ (red), $5\times 10^4~\cc$ (green), and $1\times 10^5~\cc$ (blue).
The horizontal dashed line shows the Eddington accretion for a 10\% of the radiation efficiency.

For the lowest density of $n_{\infty} = 1\times 10^4~\cc$ (Model 1e5M14N0), the accretion rate behaves 
episodically due to radiative heating (black curve in Fig. \ref{t_md_1e5M}).
The physical origin of the oscillation is explained in what follows.
In Fig. \ref{prof} (left panel), we present the radial structure of the gas density, temperature, and neutral fraction
at different three epochs in an oscillation period.
For this case, radiation associated with BH accretion propagates outward, and the gas outside the Bondi radius 
is ionized and heated up ($r_{\rm H_{II}}>r_{\rm B}$, see phase I).
Inside the ionized region, ionized gas within a new sonic radius at $\simeq 0.1~r_{\rm B}$ for the hot gas with $T\sim 10^5~\K$ 
can accrete onto the central region, while gas outside the radius flows outwards.
As a result, a density cavity forms within the ionized region, where the outward and inward gas pressure forces are balanced.
When the ionized gas is depleted from the ionized region, 
a density bump forms inside the ionization front because pressure inside decreases (phase II).
This density bump provides a positive pressure gradient ($\partial p/\partial r > 0$) and accelerate gas accretion 
($\dot{m} \approx 7.46$, phase III).
This episodic behavior has been studied in detail in previous studies
\citep[e.g.][]{Ciotti_Ostriker_2001,Milosavljevic_2009a,Park_Ricotti_2011,Park_Ricotti_2012}.
The time-averaged accretion rate results in as small as $\langle\dot{M}\rangle \approx 1.6~\dot{M}_{\rm Edd}$.

With the highest ambient density ($n_{\infty}=10^5~\cc$),
the episodic accretion behavior ceases unlike the lowest density case.
Instead, the accretion rate has a big jump to a very high value (red, green and blue curves in Fig. \ref{t_md_1e5M}).
Open circles indicate the epochs when transitions to super-Eddington accretion occur.
In Fig. \ref{prof} (right panel), we present the radial structure of the gas density, temperature, and neutral fraction
for the highest density (Model 1e5M15N0) at different three epochs of 
$t/t_{\rm dyn} = 0.334$ (phase 1), $1.25$ (phase 2), and $1.38$ (phase 3).
At the beginning, an ionized region forms and the gas is heated up to $T \sim 10^5~{\rm K}$ as in the lowest density case.
However, because of the higher density, the size of the ionized region never becomes larger than the Bondi radius ($r_{\rm H_{II}}<r_{\rm B}$). 
As a result of this, a dense shell forms at $r_{\rm H_{II}} \lesssim r \lesssim r_{\rm B}$ and pushes the ionized gas inward (phase 2).
During the transition, the ionized region shrinks and disappears because of efficient radiative recombination (phase 3).
The accretion rate jumps dramatically because the dense shell collapses and supplies a large mount of gas.
Thus, the accretion flow settles down to an isothermal Bondi accretion solution with $T\approx 8000~\K$ (blue dotted line).

For the intermediate values at $10^4~\cc < n_\infty < 10^5~\cc$,
the accretion rate begins to rise drastically at the transition epochs where a neutral shell
infalls into the center as shown in the previous case.
However, the accretion rate oscillates in short-time durations without settling to steady states 
as shown by red curve because in the burst phases, radiation force slightly exceeds ram pressure 
of neutral gas inflows and a tiny ionized region forms transiently.
As a result, the time-averaged accretion rate is as high as 
$\langle \dot{m}\rangle \simeq 2.9\times 10^3$.

We note that this episodic behavior seems a numerical artifact.
As discussed in \S \ref{sec:stable}, if all the radiation from 
the nuclear accretion disk was injected at a radius much larger than the true location of photosphere 
(i.e., $r_{\rm min}>r_{\rm ph}$), ram pressure of the inflow at the ionization front could be significantly 
underestimated and become weaker than radiation force.
Furthermore, the luminosity of ionizing photons injected at $r_{\rm min}$ would be overestimated
because the true spectrum after the transition would be softer than the slim-disk one.
Thus, the two effects are expected to cease the numerical artifact.
We demonstrate this for the case of $n_\infty =5\times 10^4~\cc$ (green curve)\footnote{
Numerical simulations setting a smaller value of $r_{\rm min}$ require a long computational time 
until the accretion flow reaches the final steady state.
Instead of this treatment, we decide to replace the radiation spectrum by a more realistic 
one with a lower mean photon energy.}, 
by replacing the spectrum with a dilute blackbody spectrum with an effective temperature
of $T_{\rm eff} = (L/4\pi R_{\rm tr}^2 \sigma_{\rm SB})^{1/4}$, where 
$\sigma_{\rm SB}$ is the Stefan-Boltzmann constant, and $r_{\rm ph}\approx R_{\rm tr}$ is approximated.
Note that this is a conservative treatment because the trapping radius is always located inside the photosphere, 
and the effective temperature measured with $R_{\rm tr}$ is higher than the true one.
As a result, we find that a super-Eddington transition stably proceeds and the accretion flow 
approaches an isothermal Bondi profile.

\begin{figure*}
\begin{center}
\includegraphics[width=16.5cm]{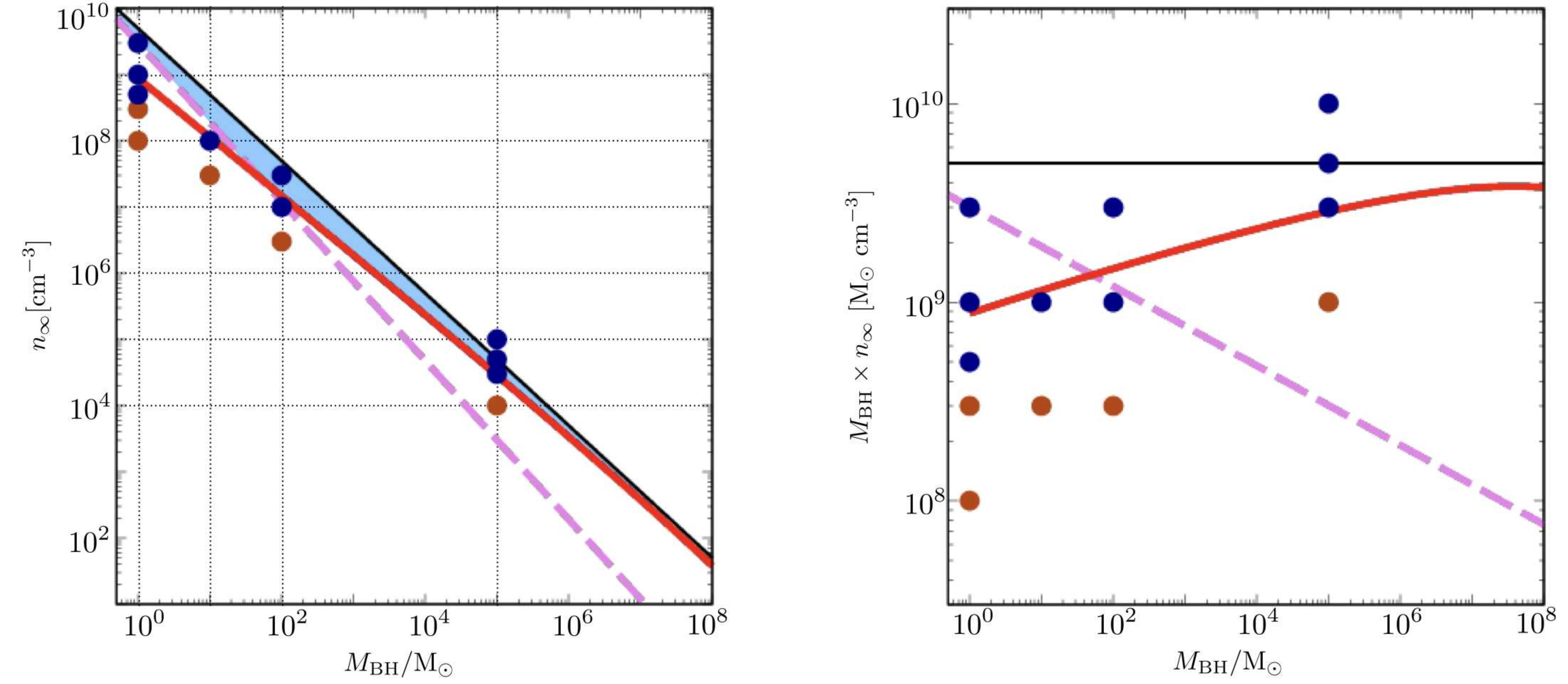}
\end{center}
\caption{
Summary of the results for different values of the BH mass $M_{\rm BH}$ and 
ambient gas density $n_{\infty}$ for isotropic radiation.
Each circle symbol indicates whether the transition to super-Eddington accretion occurs (blue) 
or the accretion rate behaves episodically (orange).
The transition criterion for disk spectra and a single PL spectrum ($L_\nu \propto \nu^{-1.5}$) are 
shown by red and black curves, respectively.
The dashed line presents the stability conditions of rapid accretion after the transition, i.e., inward ram pressure of the inflow overcomes outward force
caused by radiation with $L>L_{\rm Edd}$ (see \S\ref{sec:stable}).
The shaded region in the left panel denotes the parameter space 
where the transition criterion is alleviated from that for the power-law spectrum
and the subsequent super-Eddington accretion is stable.
}
\label{survey}
\end{figure*}

In Fig. \ref{survey}, we summarize our results for different values of $M_{\rm BH}$ and $n_{\infty}$ 
under isotropic radiation with spectra associated with accretion disks.
Each circle symbol indicates whether the transition to super-Eddington accretion occurs (blue) or the accretion rate behaves 
episodically without the transition (orange).
For the latter cases, we follow the simulations over $t>5~t_{\rm dyn}$, which is long enough to confirm the result.
The red solid (black solid) line presents the transition criterion under the disk (single PL) spectrum 
(see \S\ref{sec:ANA} for derivation).
The critical density for the transition $n_{\rm crit}$ is significantly reduced for lower BH masses with disk spectra.
In fact, we find that $n_{\rm crit,D}(M_{\rm BH})\simeq 0.1~n_{\rm crit,PL}(M_{\rm BH})$ for $M_{\rm BH}\lesssim 10^2~\msun$,
and $n_{\rm crit,D}(M_{\rm BH})\simeq 0.6~n_{\rm crit,PL}(M_{\rm BH})$ for $M_{\rm BH}\simeq 10^5~\msun$.

\subsection{Analytic arguments: the transition criterion}
\label{sec:ANA}

\begin{figure}
\begin{center}
\includegraphics[width=8.5cm]{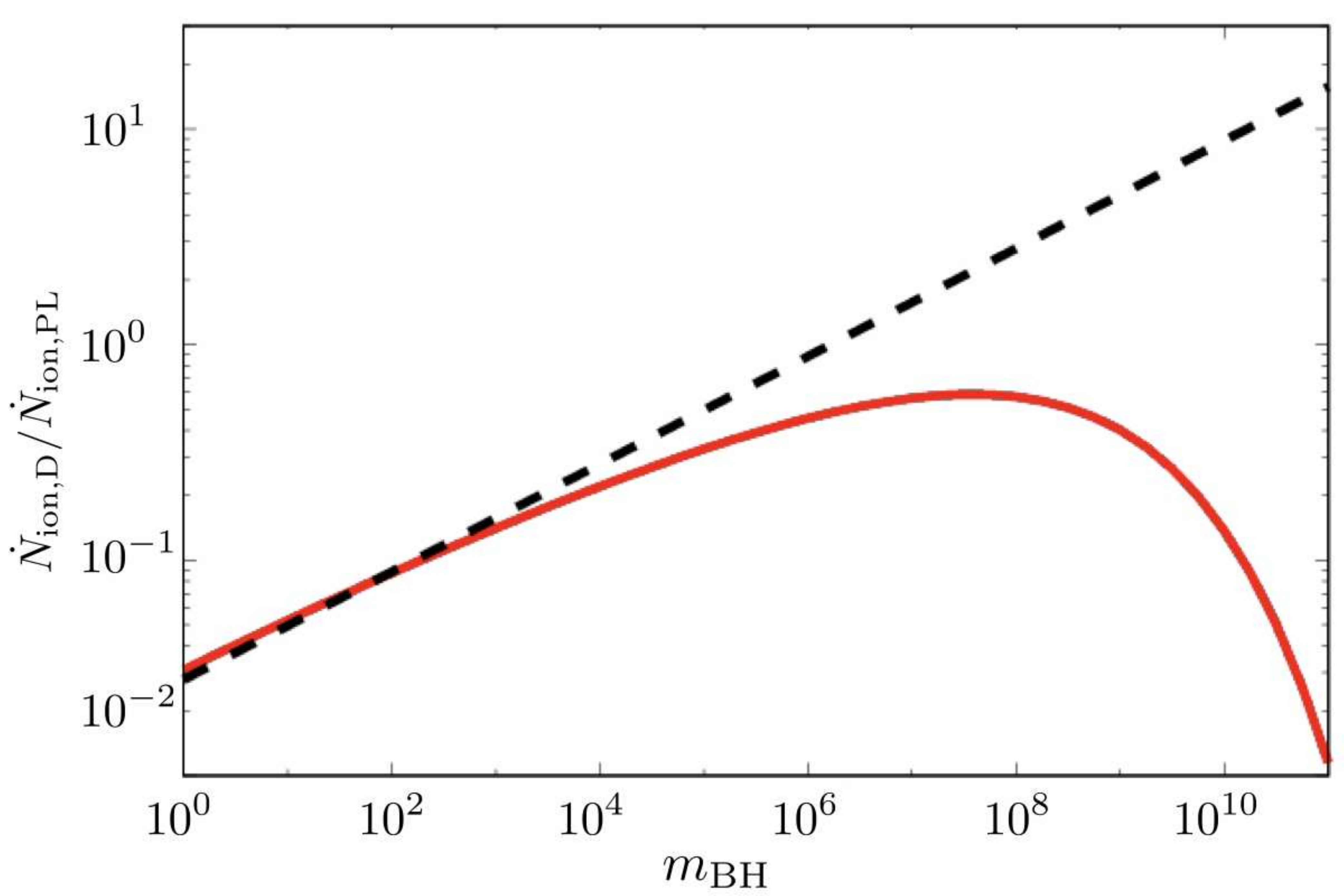}
\end{center}
\vspace{-2mm}
\caption{
The ratio $\dot{N}_{\rm ion,D} / \dot{N}_{\rm ion,PL}$ as a function of $m_{\rm BH}$ where $\dot{m}=10$ is set.
The red solid curve (black dashed line) presents the value calculated numerically (analytically, see Eq. \ref{xi_def}).
For higher BH masses, 
the analytic formula (Eq. \ref{xi_def}) is no longer valid because the peak frequency of the spectrum becomes comparable to the ionization threshold energy.
}
\label{xi}
\end{figure}

We here give a simple analytic argument for the conditions required for super-Eddington accretion,
taking into account of the radiation spectral effect.
As discussed in \cite{inayoshi+16}, the transition conditions are well explained by 
the comparison of the Bondi radius and the size of the ionized region (see Eqs. \ref{strm} and \ref{eq:bondi_radius}).
For disk spectra, the number rate of ionizing photons absorbed by neutral hydrogen within the ionization front is estimated as 
\begin{align}
\dot{N}_{\rm ion,D} \simeq \int_{\nu_{\rm min}}^{\infty}~{\rm d}\nu~ \dfrac{L_{\nu}^{\rm D,st}}{h\nu}
\approx  1.22 \times 10^{46} m_{\rm BH}^{5/4}\dot{m}^{3/4}~{\rm s^{-1}},
\label{ndot_disk}
\end{align}
where $\dot{N}_{\rm ion,D}$ is the ionizing photon number flux (in units of s$^{-1}$) and
$L_{\nu}^{\rm D,st}$ is the specific luminosity of radiation produced from a standard accretion disk
because the BH accretion rate is sub-Eddington value ($\dot{m}\simeq 10$) before the transition.
In the last expression, we neglect the contribution from ionizing photons with $\nu \geq \nu_{\rm peak}
=10^{17} m_{\rm BH}^{-1/4}\dot{m}^{1/4}~{\rm Hz}$
\footnote{Ionizing photons at frequencies of $\nu \ga 10^{17}~{\rm Hz}$ are hardly absorbed by neutral gas 
even outside the ionization front because of the steep frequency dependence of absorption cross section of neutral hydrogen ($\sigma_{\rm H}\propto \nu^{-3}$).}.
Note that this approximation causes at most $20-40~\%$ differences from the numerically integrated values for lower BH masses
with $M_{\rm BH}\la 10^5~\msun$.
Therefore, we obtain the ratio of the two radii
\begin{eqnarray}
r_{\rm H_{II}}/r_{\rm B} &\propto& (\dot{N}_{\rm ion,D})^{1/3} n_{\infty}^{-2/3}M_{\rm BH}^{-1} \nonumber \\
 & \propto& (M_{\rm BH} ^{7/8}\times n_{\infty})^{-2/3},
\end{eqnarray}
which nicely agrees to the transition criterion shown in Fig. \ref{survey} (red).
When the mean photon energy of the radiation spectrum is harder,
the ionizing photon number flux becomes smaller
for a given bolometric luminosity.
Since the intrinsic radiation spectrum is harder for the lower BH mass,
the photon absorption rate $\dot{N}_{\rm ion,D}$ becomes lower, and 
the size of the ionized region becomes relatively smaller.
Therefore, the transition to super-Eddington accretion is more likely to occur.

It is worthy comparing the transition criteria for disk spectra to those for PL spectra.
Since the ionizing photon number flux for a PL spectrum with $\alpha=1.5$ is estimated as 
\begin{equation}
\dot{N}_{\rm ion,PL} \approx 2.46 \times 10^{47} ~m_{\rm BH}\dot{m}~{\rm s^{-1}},
\end{equation}
where $L/L_{\rm Edd}\simeq 0.1~\dot{m}$ at $\dot{m}\la 20$, therefore we obtain the ratio of the two photon number fluxes
\begin{equation}
\dfrac{\dot{N}_{\rm ion,D}}{\dot{N}_{\rm ion,PL}} = 4.96 \times 10^{-2} ~m_{\rm BH}^{1/4}\dot{m}^{-1/4}.
\label{xi_def}
\end{equation}
Note that the above scaling relation is valid for lower BH masses with $M_{\rm BH}\la 10^5~\msun$ (see black dashed in Fig. \ref{xi}).
Since the peak energy  for higher BH masses becomes as low as the ionization threshold energy,
the ionization photon number sharply drops for higher BH masses (see red curve Fig. \ref{xi}).
and thus the radiation feedback effect is significantly reduced.


\begin{figure}
\begin{center}
\includegraphics[width=8.5cm]{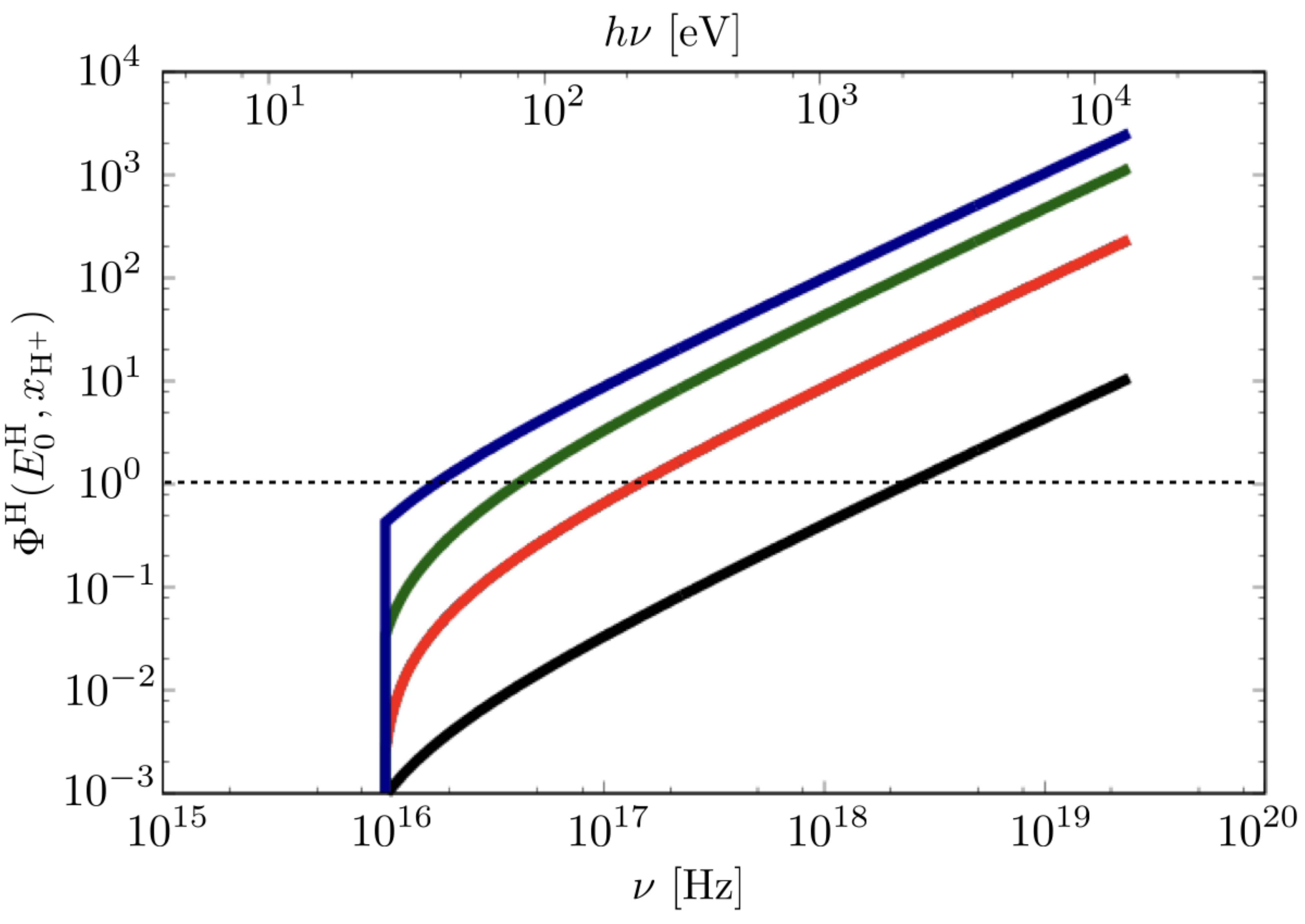}
\end{center}
\vspace{-4mm}
\caption{
The fraction of secondary ionization of H atoms per primary electron of energy $E_{0,\nu}^{i}= h\nu - I^{\rm H}$
with $x_{\rm H^{+}}=10^{-3}$ (blue), $10^{-1}$ (green), 0.4 (red), and 1.0 (black).
}
\label{slsi}
\end{figure}

We briefly mention the effects of secondary ionization on our results.
In Fig. \ref{slsi}, we show the efficiency of secondary ionization of H atoms 
per electron produced by primary ionization with a energy of $E_{\rm H} = h\nu - I_{\rm H}$
for different electron fractions of $10^{-3}\leq x_{\rm H^{+}} \leq 1.0$ (from the top to the bottom).
The horizontal dashed line presents $\Phi^{\rm H}(E_{\rm H},x_{\rm H^+})=1$, above which 
secondary ionization becomes more effective than primary ionization.
Since the photons causing primary ionization are at $\nu \lesssim 10^{17}~{\rm Hz}$,
the primary electrons hardly contribute to secondary ionization until the ionization degree increases to $\sim 0.4$.
As a result, secondary ionization can enhance the ionization degree near the ionization front, but does not 
expand the size of the ionization region.

\section{Discussion}
\label{sec:discuss}

\subsection{Mean photon energy and transition criterion}

\begin{figure}
\begin{center}
\includegraphics[width=8.2cm]{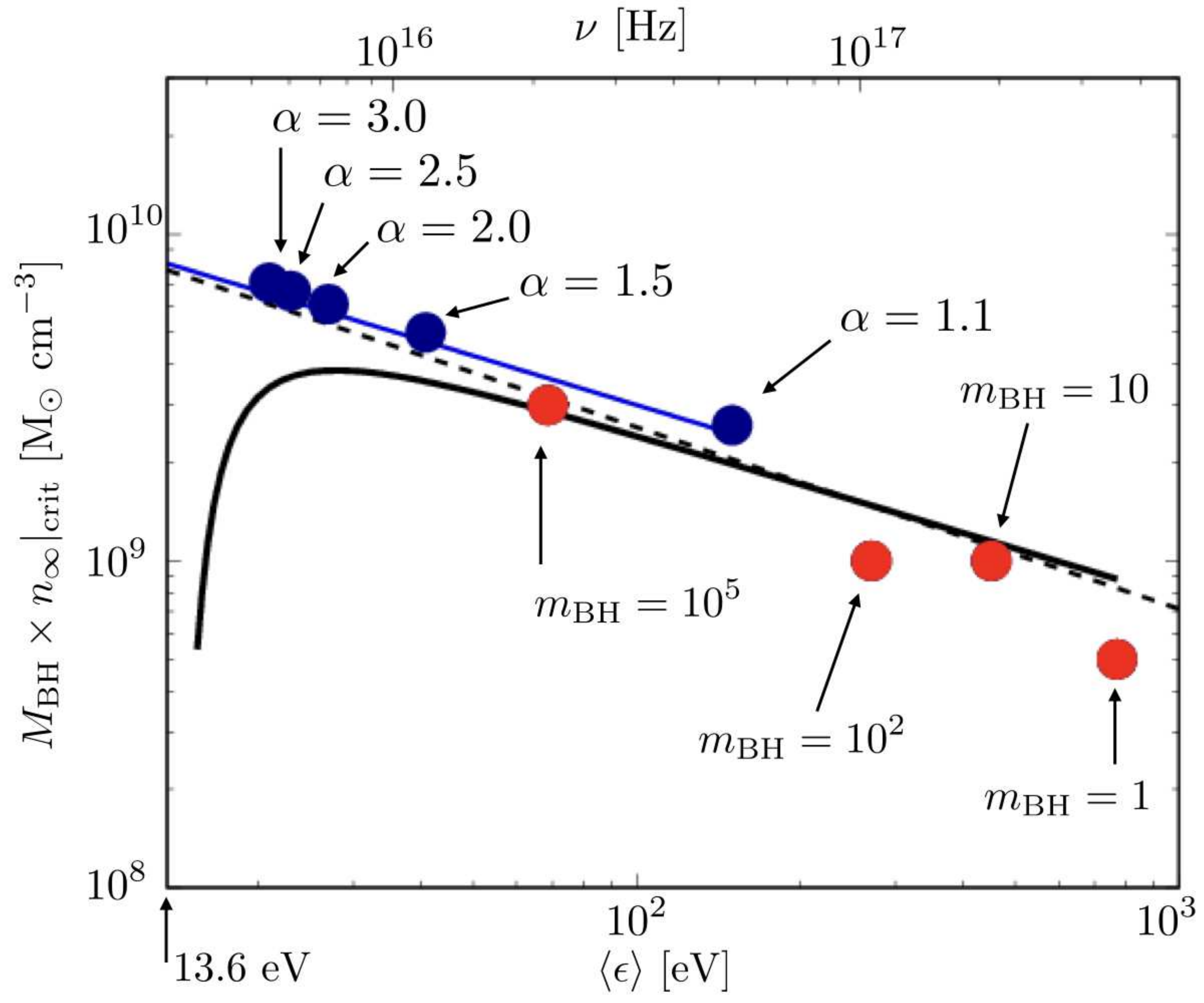}
\end{center}
\vspace{-4mm}
\caption{
The critical value of $M_{\rm BH}\times n_{\infty}$ for the transition as a function of the mean photon energy $\langle \epsilon \rangle$ for different radiation spectra.
Red circles show the numerical results for disk spectra with different BH masses.
The critical value is well explained by the dashed line: $M_{\rm BH}\times n_{\infty}\propto \langle \epsilon \rangle^{-5/9}$ 
in $70\lesssim \langle \epsilon \rangle /{\rm eV} \lesssim 800$.
Blue circles represent the analytical results for PL spectra with different PL indexes of $\alpha$, 
which are explained by blue solid line:
$M_{\rm BH}\times n_{\infty} \propto \langle \epsilon \rangle^{-1/2}$.
}
\label{matome}
\end{figure}

As described in \S\ref{sec:ANA}, the transition criterion for super-Eddington accretion depends on the shape of radiation spectra. 
Here, we generalize the criterion and rewrite the critical value of $M_{\rm BH}\times n_{\infty}$ as a function of the mean photon energy.
In Fig. \ref{matome}, we summarize our simulation results for isotropic radiation.
Red circles present the results for disk spectra for different BH masses of $M_{\rm BH}=1, ~10, ~10^2$, and $10^5~\msun$.
By setting the accretion rate to $\dot{m}=10$, we obtain the relation $\langle \epsilon \rangle = 7.7\times 10^2 ~m_{\rm BH}^{-0.23}~{\rm eV}$.
Thus, the transition criterion is expressed as 
\begin{equation}
M_{\rm BH}\times n_{\infty} \gtrsim 2.4\times 10^{9}~\msun~\cc \left(\frac{\langle \epsilon \rangle}{100~{\rm eV}} \right)^{-5/9}.
\label{fit_red}
\end{equation}
Note that this equation is no longer valid for $\langle \epsilon \rangle \lesssim 70~{\rm eV}$, 
and the critical value sharply drops at $\langle \epsilon \rangle \lesssim 20~{\rm eV}$.
In addition, blue circles present the critical values for PL spectra ($L_\nu \propto \nu^{-\alpha}$; $1.1\leq \alpha \leq 3.0$), 
where the mean photon energy is $\langle \epsilon \rangle = h\nu_{\rm min}~\alpha/(\alpha-1)$ for $\alpha > 1$,
independent of both $m_{\rm BH}$ and $\dot{m}$ \citep[see also][]{Park_Ricotti_2012}.
%
%
In the cases, the transition criterion is expressed as
\begin{equation}
M_{\rm BH}\times n_{\infty} = 3.2\times 10^{9}~\msun~\cc \left(\frac{\langle \epsilon \rangle}{100~{\rm eV}} \right)^{-1/2},
\label{fit_blue}
\end{equation}
which corresponds to the dashed line in Fig. \ref{matome}.
Note that the PL spectrum is normalized as $\eta\dot{m}L_{\rm Edd}=\int_{\nu_{\rm min}}^{\infty}{\rm d}\nu L_{\nu}$, 
where $\eta=0.1$.

Radiation spectra we observed in BH accreting systems are complex more than we considered in this paper.
In some cases, radiation spectra consist of two components: thermal emission from the nuclear disk
and non-thermal emission with PL spectra produced by Compton up-scattering in a hot corona 
\citep[e.g.,][]{haardt+1991,svensson+1994,liu+2002,liu_2003,done+2006}.
For super-Eddington accreting systems such as ULXs,
a PL component is produced by a radiation-pressure driven, hot ($\sim 10^{7.5}-10^8~\K$) outflow
where soft photons from the accretion disk are hardened by both thermal and bulk Comptonization 
\citep[e.g.,][]{kawashima+2009,kawashima+2012,Narayan+2017,kitaki+2017}.
In the following, we discuss three effects changing the mean photon energy from disk spectra.


\begin{table}
\begin{center}
\caption{The photon number flux and mean photon energy affected by Comptonization.}
\begin{tabular}{cccc}
$M_{\rm BH}/\msun$ & $\dot{N}'_{\rm abs}/\dot{N}_{\rm ion,D}$ & $\langle \epsilon' \rangle / {\rm eV}$  
&$\langle\epsilon\rangle_{\rm D} / {\rm eV}$  \\
\hline 
 $10$     & $8.4\times 10^{-2}$ & $2.0\times 10^3$ &  $4.0\times 10^2$  \\
 $10^2$ & $8.8\times 10^{-2}$ & $1.6\times 10^3$ &  $2.4\times 10^2$  \\
 $10^3$ & $9.3\times 10^{-2}$ & $1.2\times 10^3$ & $1.5\times 10^2$  \\
 $10^4$ & $0.11$ & $7.2\times 10^2$ &  $94$  \\
\hline
\end{tabular}
\label{table_ktk}
\end{center}
The number flux of ionizing photons absorbed by neutral hydrogen is calculated by 
$\dot{N}'_{\rm abs} = \int _{\nu_{\rm min}}^\infty {\rm d}\nu L'_{\nu}(1-e^{-\tau_{\nu}})/(h\nu)$,
where the spectral shape of $L'_{\nu}$ is taken from the results of \cite{kitaki+2017},
and the optical depth is estimated at the Bondi radius as $\tau_{\nu}=n_\infty r_{\rm B}\sigma_{\rm H}$. 
The mean photon energies of $\langle \epsilon' \rangle$ and $\langle \epsilon \rangle_{\rm D}$ are 
estimated by taking the data from \cite{kitaki+2017} and by assuming the disk spectra, respectively.
The accretion rate is set to $\dot{m}=10^3$.
\end{table}

\begin{figure}
\begin{center}
\includegraphics[width=8.3cm]{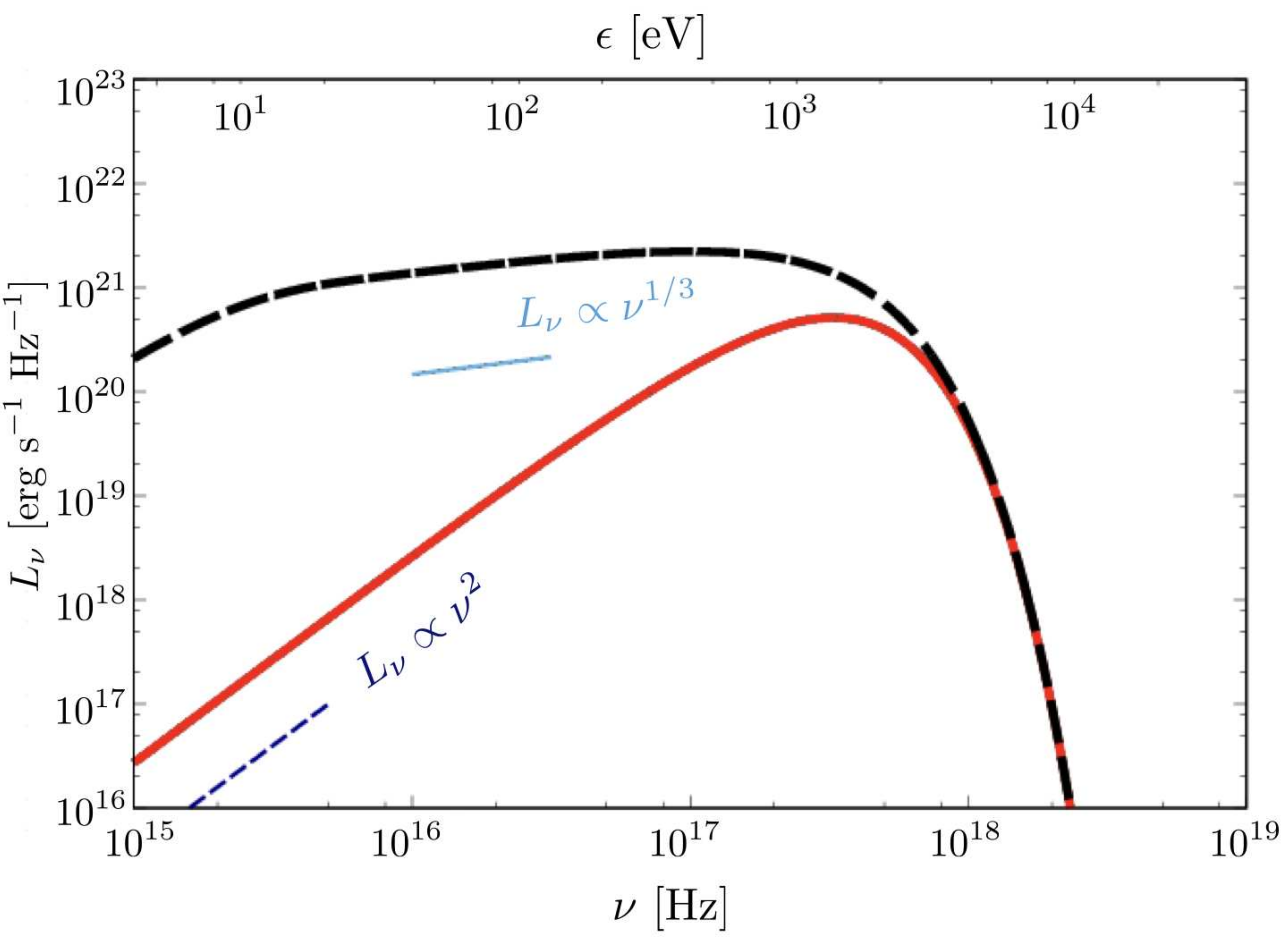}
\end{center}
\vspace{-4mm}
\caption{Dependence of radiation spectra on the choice of the disk outer edge:
$R_{\rm out}=10~r_{\rm Sch}$ (red solid) and $10^4~r_{\rm Sch}$ (black dashed).
We adopt $M_{\rm BH}=10~\msun$ and $\dot{m}=10$.
}
\label{r_out_c}
\end{figure}

Recently, \cite{kitaki+2017} studied radiation spectra of super-Eddington accretion flows onto a BH with $10\leq M_{\rm BH}/\msun \leq 10^4$ 
under a mass inflow rate of $\dot{m}\simeq 10^3$ at $R=10^3~r_{\rm Sch}$.
With Monte Carlo radiation transfer calculations, they found that a significant excess in the spectrum is produced at 
$h\nu \gtrsim {\rm a~few} ~\times$ keV due to Comptonization.
In Table \ref{table_ktk}, we summarize the number flux of ionizing photons absorbed by neutral hydrogen and mean photon energy 
estimated by taking the data from \cite{kitaki+2017} (lower panels of their Fig. 5).
Compared to the cases assuming disk spectra, the mean energies are boosted by a factor of $5-8$.
Since such hard X-rays with $h\nu \gtrsim {\rm keV}$ are hardly absorbed even by neutral hydrogen, 
the numbers of absorbed photons are reduced by one order of magnitude ($\dot{N}'_{\rm abs}/\dot{N}_{\rm ion,D} \simeq 0.1$).
Therefore, the transition criterion is alleviated by a factor of $\approx 3$.


The size of the nuclear accretion disk would affect the feedback efficiency because ionizing photons 
with $13.6~{\rm eV} \la h\nu \la 1~{\rm keV}$ are produced from larger disk radii.
In Fig. \ref{r_out_c}, we demonstrate the dependence of radiation spectra on the choice of the disk outer edge: 
$R_{\rm out}=10^4~r_{\rm Sch}$ (black) and $10~r_{\rm Sch}$ (red) for $M_{\rm BH}=10~\msun$ and $\dot{m}=10$.
For $R_{\rm out}=10~R_{\rm Sch}$, the spectrum is no longer expressed as a multi-color blackbody spectrum but by the Rayleigh-Jeans law.
Thus, the number flux of ionizing photons at $\nu \lesssim 3 \times 10^{17}~{\rm Hz}$ is significantly reduced and 
the mean photon energy increases to $\langle\epsilon\rangle \simeq 1.3~{\rm keV}$ from $\langle\epsilon\rangle \simeq 450~{\rm eV}$.
As a result, the critical value for the transition would be reduced only by a factor of $\approx 2$.
If $R_{\rm out}$ is much larger than the fiducial value,
the emission rate of photons would increase only in the energy range with $h\nu < 13.6~{\rm eV}$. 
Since these photons are less energetic to ionize the ambient gas, 
the radiative feedback would not be enhanced, and the critical value for transitions would not be changed\footnote{
The choice of $R_{\rm out}$ depends on the angular momentum of inflowing gas from the Bondi radius.
When the gas is optically thin to Ly$\alpha$ lines, a quasi-hydrostatic dense torus with a constant temperature of 
$T\simeq 8000~\K$ forms around the centrifugal radius of $r_{\rm cent}~(<r_{\rm B})$.
As long as the angular momentum is so small that $r_{\rm cent} \la 0.03~r_{\rm B}$ is satisfied,
the rate of accretion driven by viscosity can be comparable to the Bondi rate \citep{sugimura_2018}.}
.


In addition, the existence of dust grains in accretion flows significantly change the spectral shape 
due to UV attenuation caused by dust absorption and thus alleviate the criterion for super-Eddington accretion significantly.
Recent work by \cite{Toyouchi+2018} has found that rapid accretion of metal-polluted gas is allowed
as long as $Z\lesssim 10^{-2}~\zsun$, because ionizing radiation from the central BH is absorbed and reemitted to infrared lights 
with lower energies ($h \nu \ll 13.6~{\rm eV}$).

\subsection{Stability condition}
\label{sec:stable}

We briefly discuss the stability of accretion flows at a very high inflow rate of $\dot{m}> 10^3$, exposed to 
intense radiation with a luminosity of $L> L_{\rm Edd}$ \citep[see also][]{Sakurai_2016}.
As shown in Fig. \ref{prof}, during the transition a dense shell of neutral gas pushes the ionization front
and a neutral region propagates inwards where a high accretion rate builds up.
Thus, when the central BH is actually fed by the inflow and produces intense radiation, 
the accreting flow would already settle in an isothermal Bondi solution with $T\approx 8000~\K$.
However, the accretion flow cannot be neutral all the way down to the BH, but is ionized at a smaller radius $r_{\rm ph}$ 
where the flow becomes optically thick to continuum absorption.
Considering ${\rm H}^-$ bound-free absorption opacity, we can estimate the location of the photosphere\footnote{
In a partially ionized region, the strong dependence of opacity on gas temperature leads to a thermal-ionization instability 
\citep[e.g.,][]{meyer+1981, kato+2008}.
Though the location of the photosphere $r_{\rm ph}$ results in time-dependent, our order-of-magnitude estimate is not
significantly changed.} as
\begin{align}
r_{\rm ph} \simeq \left( \frac{3GM_{\rm BH} \dot{M}_{\rm B}}{ 8\pi\sigma_{\rm SB}T_{\rm ph}^4} \right)^{1/3}
\simeq 7.1\times 10^{13}~{\rm cm}~
\left(\frac{\dot{m}}{10^4}\right)^{1/3} 
\left(\frac{m_{\rm BH}}{10^4}\right)^{2/3},
\label{r_ph}
\end{align}
where the photospheric temperature is set to $T_{\rm ph} = 2\times 10^4~\K$ \citep{inayoshi+16}.
Assuming that all incident radiation emitted from the photosphere is absorbed by the inflowing neutral gas and deposits momentum of $L/c$,
the inflow velocity of neutral gas is accelerated to the free-fall value $v_{\rm ff}$ and push the gas at $r_{\rm ph}$ 
with ram pressure of $\dot{M}_{\rm B}v_{\rm ff}$.
Therefore, the stability condition for super-Eddington accretion (i.e., $\dot{M}_{\rm B}v_{\rm ff}\gtrsim L/c$) is rewritten as
\begin{equation}
\frac{L}{L_{\rm Edd}} \lesssim 83 ~
\left(\frac{\dot{m}}{10^4}\right)^{5/6}
\left(\frac{m_{\rm BH}}{10^4} \right)^{1/6}.
\end{equation}
As shown in Fig. \ref{survey}, super-Eddington accretion for higher BH masses ($M_{\rm BH}\gtrsim 10^2~\msun$)
satisfy the stability condition after the transition occurs.

\begin{figure*}
\begin{center}
\includegraphics[width=15cm]{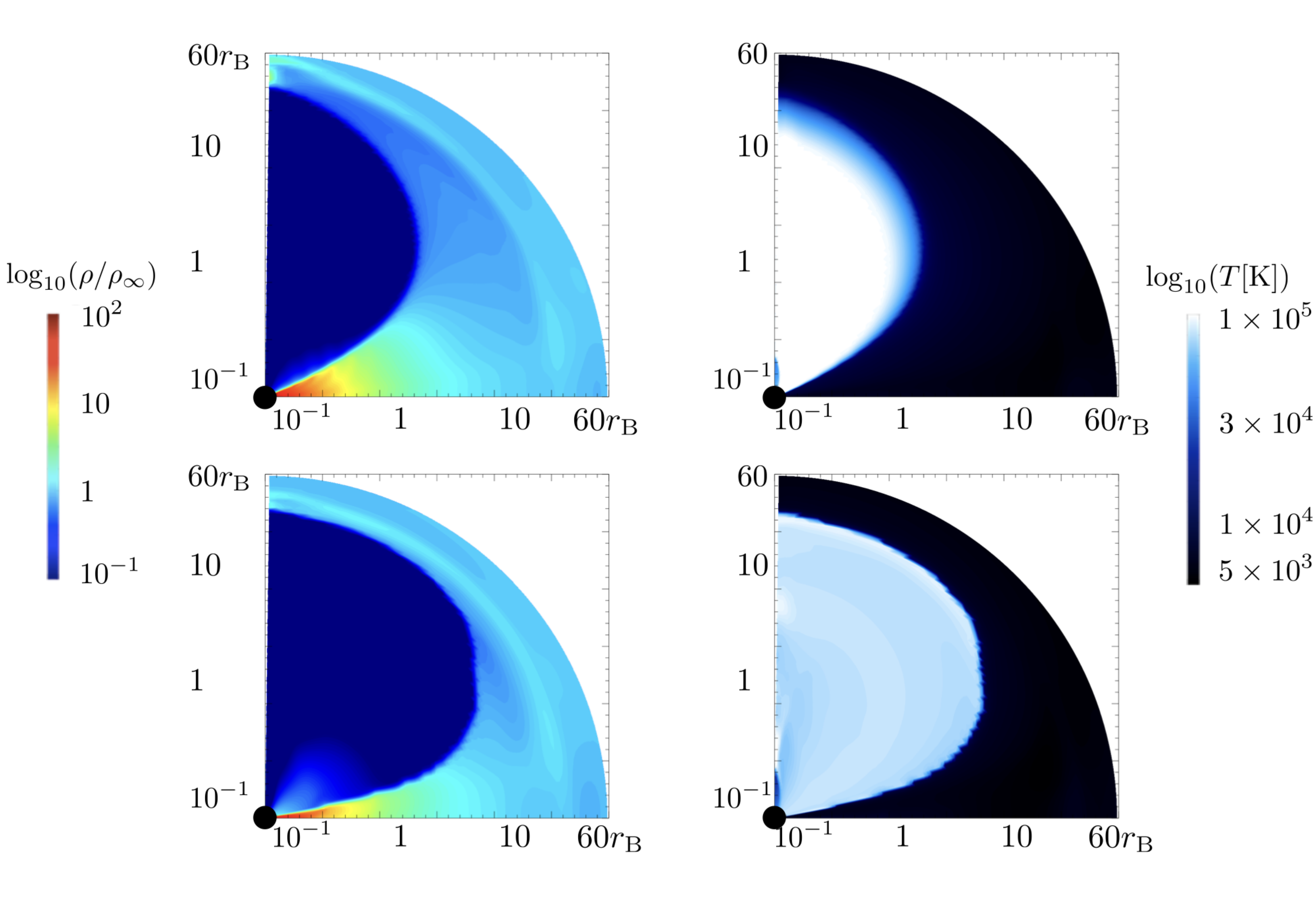}
\end{center}
\vspace{-4mm}
\caption{
Two-dimensional distribution of
the gas density (left panels) and the temperature (right panels)
under the disk spectrum (top panels) and the PL spectrum (bottom panels)
in cases with $1~\msun$, $10^9~\cc$, and $\mathcal{N}=4$.
The elapsed time is $t = 1.7 \times 10^2~{\rm yr} = 20~t_{\rm dyn}$.
}
\label{cont_2d}
\end{figure*}

\subsection{Cases with anisotropic radiation}
\label{sec:anisotropic}

We also examine cases with anisotropic disk radiation spectra.
Under anisotropic radiation with a PL spectrum, hot ionized gas expands towards the bipolar directions,
and the neutral warm gas with $T \simeq 8000~\K$ accretes through the equatorial plane at a rate of 
$\dot{M} \simeq \dot{M}_{\rm B} \sin{\Theta}$
\citep{sugimura_2017,takeo_2018},
where $\Theta$ is the half angle of the neutral region measured from the equator.
Moreover, the transition to efficient accretion where the entire region is covered by neutral gas occurs 
when $M_{\rm BH}\times n_{\infty} \gtrsim 5\times 10^{10}~\msun~\cc$ is satisfied \citep{takeo_2018}.
We mention how those features are affected by disk spectra in cases with/without transitions, respectively.

In cases without transitions, we compare a quantity of $\sin{\Theta}$ for the two cases. 
Fig. \ref{cont_2d} presents two-dimensional distribution of the gas density (left panels) and temperature (right panels)
under the disk spectrum (top) and the PL spectrum (bottom) for $M_{\rm BH}\times n_{\infty} = 10^9~\msun ~\cc$ and $\mathcal{N}=4$ 
(hereafter Model 1e0M19N4; $\dot{M}_{\rm B}/\dot{M}_{\rm Edd}=7000$).
The half angle for the disk spectrum ($\Theta_{\rm D} \simeq 19^{\circ}$) at the Bondi radius becomes twice larger  
than that for the PL spectrum ($\Theta_{\rm PL} \simeq 10^{\circ}$).
We can estimate the opening angle by solving $r_{\rm B}=r_{\rm H_{\rm II}}(\Theta)$, where
\begin{equation}
r_{\rm H_{II}}(\Theta)\approx \left[ \frac{3(\mathcal{N}+1) \dot{N}_{\rm ion} \sin^\mathcal{N} \Theta}{4\pi n^2 \alpha_{\rm B}}\right]^{1/3},
\end{equation}
and the angular dependence reflects the anisotropic radiation flux given in Eq. (\ref{frad_in}).
Therefore, the ratio of the half opening angle for the two cases is evaluated as 
\begin{equation}
\dfrac{\sin{\Theta_{\rm D}}}{\sin{\Theta_{\rm PL}}} 
= \left(\dfrac{\dot{N}_{\rm ion, D}}{\dot{N}_{\rm ion,PL}}\right)^{-1/\mathcal{N}}.
\label{theta_HI}
\end{equation}
For the disk spectrum, the ionizing photon number flux is given by 
\begin{equation}
\dot{N}_{\rm ion, D} \simeq 
\int_{\nu_{\rm tr}}^{\nu_{\rm in}}{\rm d}\nu~\dfrac{L_{\nu}^{\rm D,slim}}{h\nu} \simeq 3.0\times 10^{48}~m_{\rm BH}^{5/4}~{\rm s}^{-1},
\end{equation}
where we approximately estimate $\dot{N}_{\rm ion, D}$ taking account of the slim disk component because the accretion rate 
is as high as $\dot{m}\sim O(10^3)$.
For a single power-law spectrum, the photon flux is calculated as
\begin{equation}
\dot{N}_{\rm ion,PL} = 4.90 \times 10^{48} m_{\rm BH}\left[1+{\rm ln}\left(\frac{\dot{m}}{20}\right) \right]~{\rm s^{-1}}.
\end{equation}
where the luminosity is estimated by Eq. (\ref{eq:Lmdot}).
Therefore, we obtain the analytical expression of the ratio of the half opening angles for $\mathcal{N}=4$
\begin{equation}
\dfrac{\sin{\Theta_{\rm D}}}{\sin{\Theta_{\rm PL}}} 
\simeq 1.13~m_{\rm BH}^{-1/16}\left[1+\ln \left(\frac{\dot{m}}{20}\right)\right]^{1/4}.
\label{theta_HI2}
\end{equation}
This analytical expression agrees with the numerical results within errors of $\lesssim 10 \%$.

In cases with the transition to the wholly neutral phase, the critical conditions can be derived by equating 
$r_{\rm H_{II}}/r_{\rm B}$ at poles towards which the radiation flux is collimated,
\begin{equation}
\frac{M_{\rm BH} \times n_{\infty}}{10^9~\msun ~\cc} \gtrsim \sqrt{\mathcal{N}+1}
\begin{cases}
7.1 \left[1+{\rm ln}\left(\dot{m}/20\right)\right]^{1/2} &{\rm (PL)},\vspace{1mm}\\
8.0 \left(m_{\rm BH}/10\right)^{1/8} &{\rm (Disk)}
  \end{cases}
\end{equation}
We note that the criterion for $\mathcal{N}=4$ agrees with the numerical result shown in \cite{takeo_2018}.

\subsection{Comparison with previous studies}

Finally, we discuss the environmental conditions where the super-Eddington transition takes place.
For stellar-mass BHs originated from gravitational collapse of massive Pop III stars, 
various negative feedback effects (stellar irradiation, energetic supernovae explosions, and BH radiation feedback)
evacuate gas from shallow gravitational potential well of the host dark-matter halo and thus
likely prevent the remnant BHs from accreting the ambient gas at super-Eddington rates
\citep[e.g.,][]{kitayama+2004, Kitayama+2005, Johnson+2007, Alvarez_2009}. 
Even when a seed BH is embedded a relatively massive halo, so-called atomic-cooling halo with a virial temperature of
$\sim 10^4~\K$, where gas would be tightly bound in the halo potential well,  
BH feedback and energetic supernovae would expel the gas at the vicinity of the BH and shut the BH growth off
\citep[e.g.,][]{Johnson+2011, Aykutalp+2013, Dubois+2012, Dubois+2015, Prieto+2016, Smidt+2018, Latif+2018}. 
Although some studies mentioned above marginally resolve the Bondi radius for neutral warm gas with $T\sim 8000~\K$,
their prescriptions for energy and/or momentum feedback injected in unresolved regions still remain uncertain;
namely, the density threshold above which the gas turns into stars is much smaller than the critical value 
for super-Eddington transitions for a heavy seed BH with $M_{\rm BH}=10^5~\msun$. 
Further studies using high-resolution cosmological simulations will be left in future.
On the other hand, stellar-mass BHs embedded in the central gas-rich region of an atomic-cooling halo 
might grow at super-Eddington rates \citep[e.g.,][]{Lupi_2016, Ryu_2016}.

In this paper, we have explored several cases with different gas densities surrounding a BH.
In reality, however, the boundary conditions would be set by external influences (e.g. rapid major mergers 
with other haloes) associated with cosmological large-scale structures.
In the recent decade, large-scale cosmological simulations studying galaxy formation and evolution
have been carried out intensively, e.g., FIRE \citep[e.g.,][]{Hopkins+2014}, and 
Illustris \citep[e.g.,][]{Vogelsberger+2014a} simulation.  
Some simulation studies \citep[e.g., ][]{Habouzit_2017,DiMatteo+2017} focused on the early epoch of the Universe
and investigated the relationships between growth of high-$z$ BHs and the properties of their host halos. 
As a sub-grid model to characterize rapid growth of BHs, the transition conditions would be applicable 
for those cosmological simulations which do not resolve the Bondi radius of the BHs.

\section{Summary and Conclusions}
\label{sec:conc}

We investigate the properties of accretion flows onto a BH with a mass of $M_{\rm BH}$
embedded in an initially uniform gas cloud with a density of $n_{\infty}$
in order to study rapid growth of BHs in the early Universe.
In previous work, the conditions required for super-Eddington accretion from outside the Bondi radius were studied 
by assuming that radiation produced at the vicinity of the central BH has a single-power-law spectrum $\nu^{-\alpha}$ 
at $h\nu \geq 13.6~{\rm eV}$ ($\alpha \sim 1.5$).
However, radiation spectra surely depends on the BH mass and accretion rate, and determine the efficiency of radiative feedback.
Here, we perform two-dimensional multi-frequency radiation hydrodynamical simulations taking into account more realistic 
radiation spectra associated with the properties of nuclear accretion disks.
We find that the critical density of gas surrounding the BH, above which a transitions to super-Eddington accretion occurs,
is alleviated for a wide range of masses of seed BHs ($10\lesssim M_{\rm BH}/\msun \lesssim 10^6$)
because photoionization for accretion disk spectra are less efficient 
than those for single-power-law spectra with $1\lesssim \alpha \lesssim 3$.
For disk spectra, the transition to super-Eddington is more likely to occur for lower BH masses 
because the radiation spectra become too hard to ionize the gas.
Even when accretion flows are exposed to anisotropic radiation, the effect due to radiation spectra shrinks the ionized region
and and likely leads to the transition to a wholly neutral accretion phase.
Finally, by generalizing our simulation results, we construct a new analytical criterion required for super-Eddington accretion,
\begin{equation}
\left(\frac{M_{\rm BH}}{10^5~\msun}\right) \left(\frac{n_{\infty}}{10^4~\cc}\right) \gtrsim 2.4~ \left(\frac{\langle\epsilon\rangle}{100~{\rm eV}}\right)^{-5/9}
\end{equation}
where $\langle\epsilon\rangle$ is the mean energy of ionizing radiation from the central BH.

\section*{Acknowledgements}

We would like to thank Takaaki Kitaki for providing the spectral data of super-critical accretion flows.
We also thank Daisuke Toyouchi for fruitful discussions.
This work is partially supported by Japan Society for the Promotion of Science Grant-in-Aid
for Scientific Research (A) (17H01102 KO), Scientific Research (C) (16K05309 KO; 18K03710 KO; 17K0583 SM),
Scientific Research on Innovative Areas (18H04592 KO), and Young Scientists (17K14260 HRT),
and supported by the National Key R\&D Program of China (2016YFA0400702), and the 
National Science Foundation of China (11721303).
This research is also supported by the Ministry of Education, Culture, Sports, Science and Technology of Japan 
as "Priority Issue on Post-K computer" (Elucidation of the Fundamental Laws and Evolution of the Universe) and JICFuS.
Numerical computations were carried out on Cray XC30 and XC50 at Center for Computational Astrophysics, 
National Astronomical Observatory of Japan.





\bibliography{ref.bib}







\bsp	
\label{lastpage}
\end{document}